\documentclass[sn-mathphys-num,pdflatex]{sn-jnl}

\usepackage[T1]{fontenc}
\usepackage{lmodern}
\usepackage{graphicx}%
\usepackage[inline]{enumitem}%
\usepackage{amsmath,amssymb,amsfonts}%
\usepackage{textcomp,gensymb}%
\usepackage{microtype}%

\begin{document}

\title{On the determination of the interaction time of GeV neutrinos in large argon gas TPCs}

\author[a]{A.~Sa\'a-Hern\'andez}
\author*[a]{D.~Gonz\'alez-D\'iaz}
\email{diego.gonzalez.diaz@usc.es}
\author*[b]{J.~Mart\'in-Albo}
\email{justo.martin-albo@ific.uv.es}
\author[b]{M.~Tuzi}
\author[a,b]{P.~Amedo}
\author[c]{J.~Baldonedo}
\author[b]{C.~Ben\'itez}
\author[d]{S.~Bounasser}
\author[c]{E.~Casarejos}
\author[c]{J.~Collazo}
\author[a]{A.~Fernández-Prieto}
\author[a]{D.~J.~Fernández-Posada}
\author[b]{R.~Hafeji}
\author[a]{S.~Leardini}
\author[a]{D.~Rodas-Rodríguez}
\author[c]{A.L.~Saborido}
\author[c]{A.~Segade}
\author[a]{A.~Slater}

\affil[a]{\orgdiv{Instituto Galego de F\' isica de Altas Enerx\' ias (IGFAE)}, \orgname{Universidade de Santiago de Compostela}, \orgaddress{\street{Rúa de Xoaquín Díaz de Rábago, s/n}, \postcode{15782} \city{Santiago de Compostela}, \country{Spain}}}

\affil[b]{\orgdiv{Instituto de F\' isica Corpuscular (IFIC)}, \orgname{CSIC \& Universitat de València}, \orgaddress{\street{Calle Catedrático José Beltrán, 2}, \postcode{46980} \city{Paterna, Valencia}, \country{Spain}}}

\affil[c]{\orgdiv{Centro de Investigación en Tecnologías, Energías y Procesos Industriales (CINTECX)}, \orgname{Universidade de Vigo}, \orgaddress{\postcode{36310} \city{Vigo, Pontevedra} \country{Spain}}}

\affil[d]{\orgdiv{Department of Physics}, \orgname{Harvard University}, \orgaddress{\street{17 Oxford St}, \city{Cambridge}, \state{MA} \postcode{02138}, \country{USA}}}

\abstract{Next-generation megawatt-scale neutrino beams open the way to studying neutrino-nucleus scattering using gaseous targets for the first time. This represents an opportunity to improve the knowledge of neutrino cross sections in the energy region between hundreds of MeV and a few GeV, of interest for the upcoming generation of long-baseline neutrino oscillation experiments. The challenge is to accurately track and (especially) time the particles produced in neutrino interactions in large and seamless volumes down to few-MeV energies. We propose to accomplish this through an optically-read time projection chamber (TPC) filled with high-pressure argon and equipped with both tracking and timing functions. In this work, we present a detailed study of the time-tagging capabilities of such a device, based on end-to-end optical simulations that include the effect of photon propagation, photosensor response, dark count rate and pulse reconstruction. We show that the neutrino interaction time can be reconstructed from the primary scintillation signal with a precision in the range of 1--2.5~ns ($\sigma$) for point-like deposits with energies down to 5~MeV. A similar response is observed for minimum-ionizing particle tracks extending over lengths of a few meters. A discussion on previous limitations towards such a detection technology, and how they can be realistically overcome in the near future thanks to recent developments in the field, is presented. The performance demonstrated in our analysis seems to be well within reach of next-generation neutrino-oscillation experiments, through the instrumentation of the proposed TPC with conventional reflective materials and a silicon photomultiplier array behind a transparent cathode.}


\maketitle

\section{Introduction} 
\label{sec:Introduction}
The \emph{time projection chamber} (TPC) has been at the forefront of particle physics research for the last four decades, finding numerous applications in areas ranging from heavy-ion physics to rare-event searches \cite{Nygren:1974nfi, Hilke:2010zz, Nygren:2018sjx}. In neutrino physics, the liquid argon TPC (LArTPC), originally proposed in the late 1970s \cite{Rubbia:1977zz}, is now the cornerstone detector technology of the short- \cite{Acciarri:2015bmn} and long-baseline \cite{DUNE:2020lwj} neutrino oscillation programs under development in the United States. More recently, magnetized gaseous argon TPCs (GArTPCs) have also been proposed as neutrino detectors \cite{Andreopoulos:2284748, DUNE:2022yni}. Their combination of high-resolution tracking, ability to reconstruct short hadron tracks down to few MeV, and powerful particle identification through d$E/$d$x$ would help improve our knowledge of neutrino-nucleus scattering in the few-GeV region, reducing systematic errors in the new generation of neutrino oscillation experiments.

A key aspect of the design of a GArTPC is that, in contrast to the liquid phase, high-resolution tracking in pure argon gas is not viable due to
\begin{enumerate*}[label=(\roman*)]
    \item high electron diffusion, and
    \item low primary ionization per mm combined with a very low avalanche amplification factor (around $15$--$30~\mathrm{e^-/e^-}$ \cite{Cantini:2014xza,Tesi:2023,Olano:2023}).
\end{enumerate*}
Illustratively, at 10~bar and for a typical drift field of 400~V/cm, the ionization spread resulting from a 5~m drift would be $\sigma_\mathrm{T\thinspace(L)} \sim 2.2\thinspace(0.8)$~cm~\cite{Pack:1992}.\footnote{The subscripts T and L stand, respectively, for the charge spread \emph{transverse} (perpendicular) or \emph{longitudinal} (colinear) to the drift field.} Such diffusion values are more than one order of magnitude higher than those usually found in state-of-the-art gaseous TPCs, while the gain is at least $100$ times lower (see, e.g., ref.~\cite{Alme:2010ke}). Electron diffusion can be greatly reduced through the addition of a small fraction of a molecular gas \cite{NEXT:2018qyk} (typically, at the few-percent level or above). This helps as well with the stabilization of the avalanche process \cite{Gonzalez-Diaz:2017gxo} by absorbing the very energetic scintillation photons emitted by rare gases, as well as quenching its emission \cite{Takahashi:1983}. 

\begin{figure}
\centering
\includegraphics[width=0.40\textwidth]{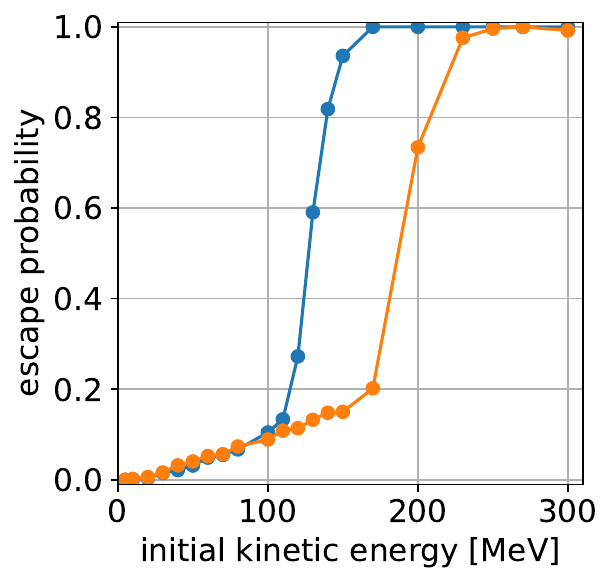}
\caption{Escape probability (estimated using Geant4 \cite{Allison:2016lfl, Allison:2006ve, GEANT4:2002zbu}) of protons from a cylindrical TPC, 5~m in diameter and 5 m long, housed in a high-pressure steel vessel filled with Ar/CF$_4$ (99/1) at 10~bar. The blue data correspond to a 1-cm-thick vessel; the orange data to a 3-cm-thick one. Protons were generated isotropically from the center of the detector.}
\label{fig:ProtonEscapeProb}
\end{figure}

The above creates an apparent technological dilemma: in a GArTPC, good tracking capabilities and usable levels of scintillation seem to be incompatible. In the LArTPCs used as neutrino detectors, the primary-scintillation signal provides the absolute timing of the recorded tracks, which is required to disambiguate their position along the drift direction. In a GArTPC, although one could in principle rely on external instrumentation to time those neutrino interactions through tracks escaping the TPC, that would leave unreconstructed the 3D position of many reactions of interest. Even in the gaseous phase, tracks of up to 100--200 MeV may range-out inside the TPC depending on the conditions (see figure~\ref{fig:ProtonEscapeProb}). In other words, the main advantage of a GArTPC, its ability to reconstruct low-energy  interactions, would be hindered by its inability to time-tag them.

An appealing path forward is to induce wavelength-shifting of the argon scintillation directly in the gas, as this process can simultaneously remove the avalanche-destabilizing vacuum-ultraviolet (VUV) photons from argon, while preserving a significant fraction of the scintillation yield. One can anticipate additional advantages like easier light collection and detection away from the VUV, or a faster time response (given the very long time constant of 3.2~{\textmu}s from the Ar$_2^*$ triplet state \cite{Amsler:2007gs}, responsible  for about 80--90\% of the argon scintillation emission \cite{Santorelli:2020fxn}). If gaseous wavelength-shifting can be accomplished through a molecular species, then the suppression of diffusion to manageable levels as well as the enhancement of the avalanche multiplication factor can be expected too.

\begin{figure}
\centering
\includegraphics[width=0.495\textwidth]{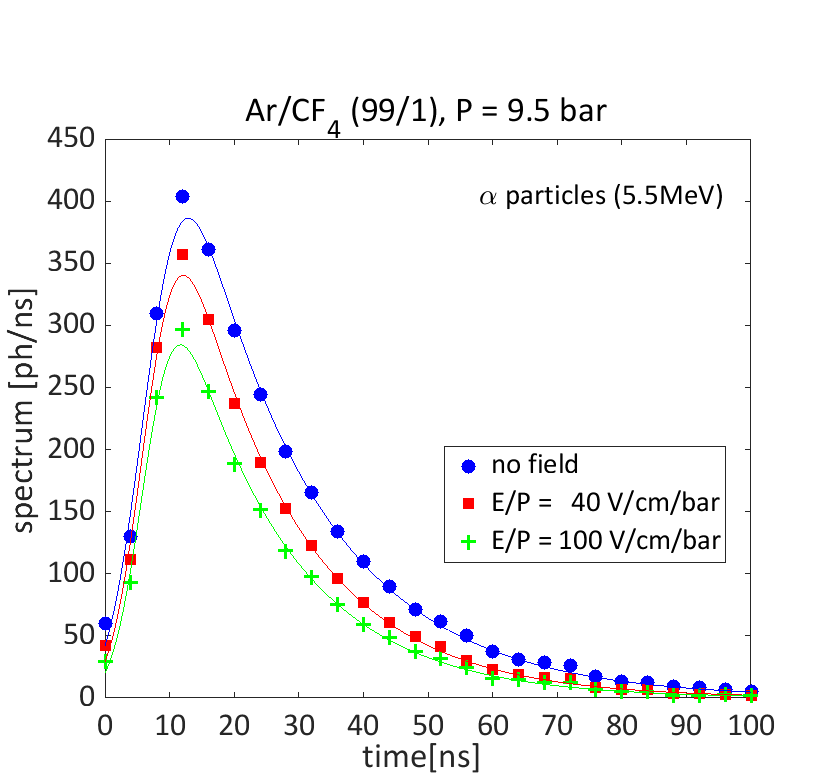}
\includegraphics[width=0.495\textwidth]{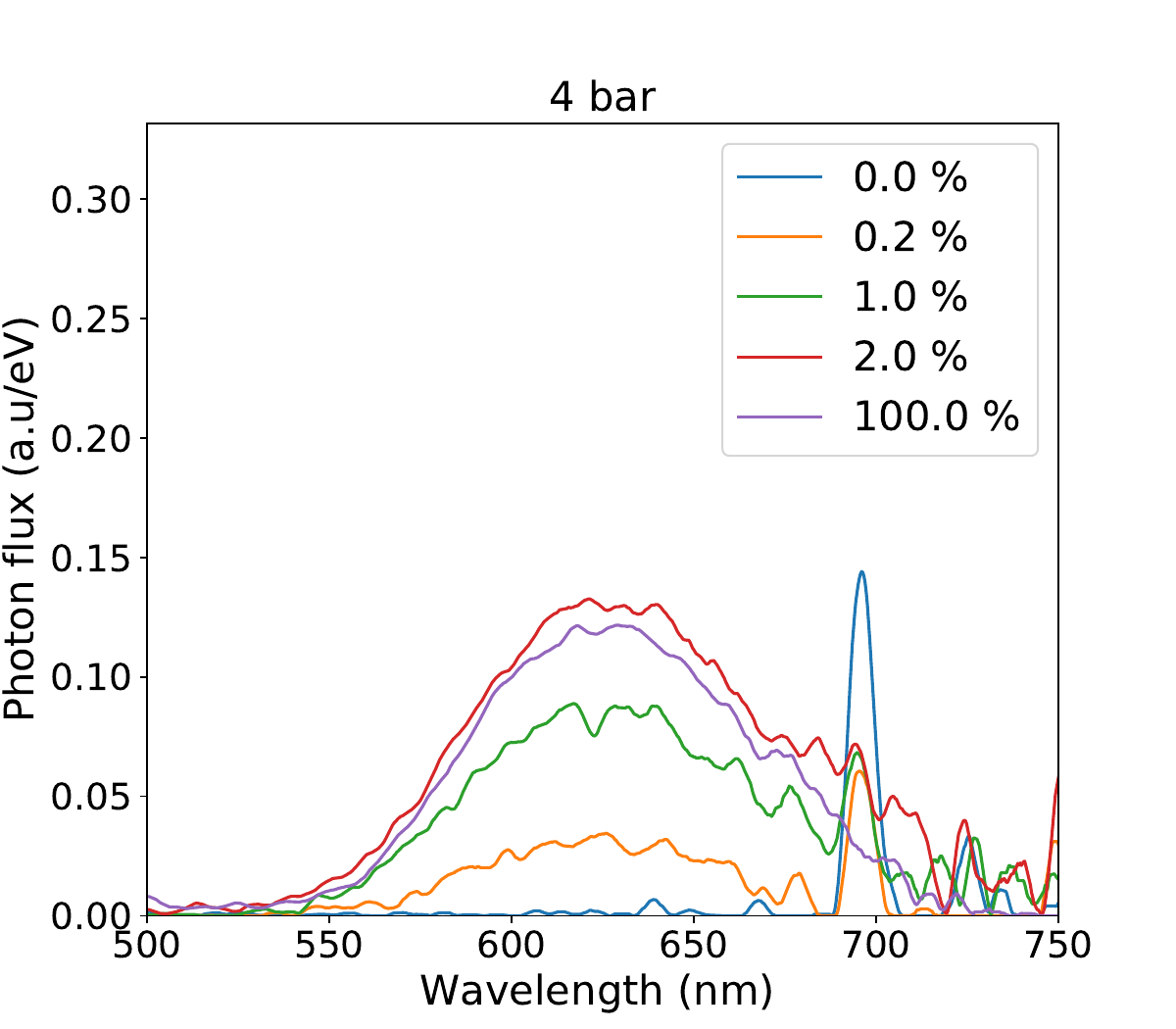}
\includegraphics[width=0.495\textwidth]{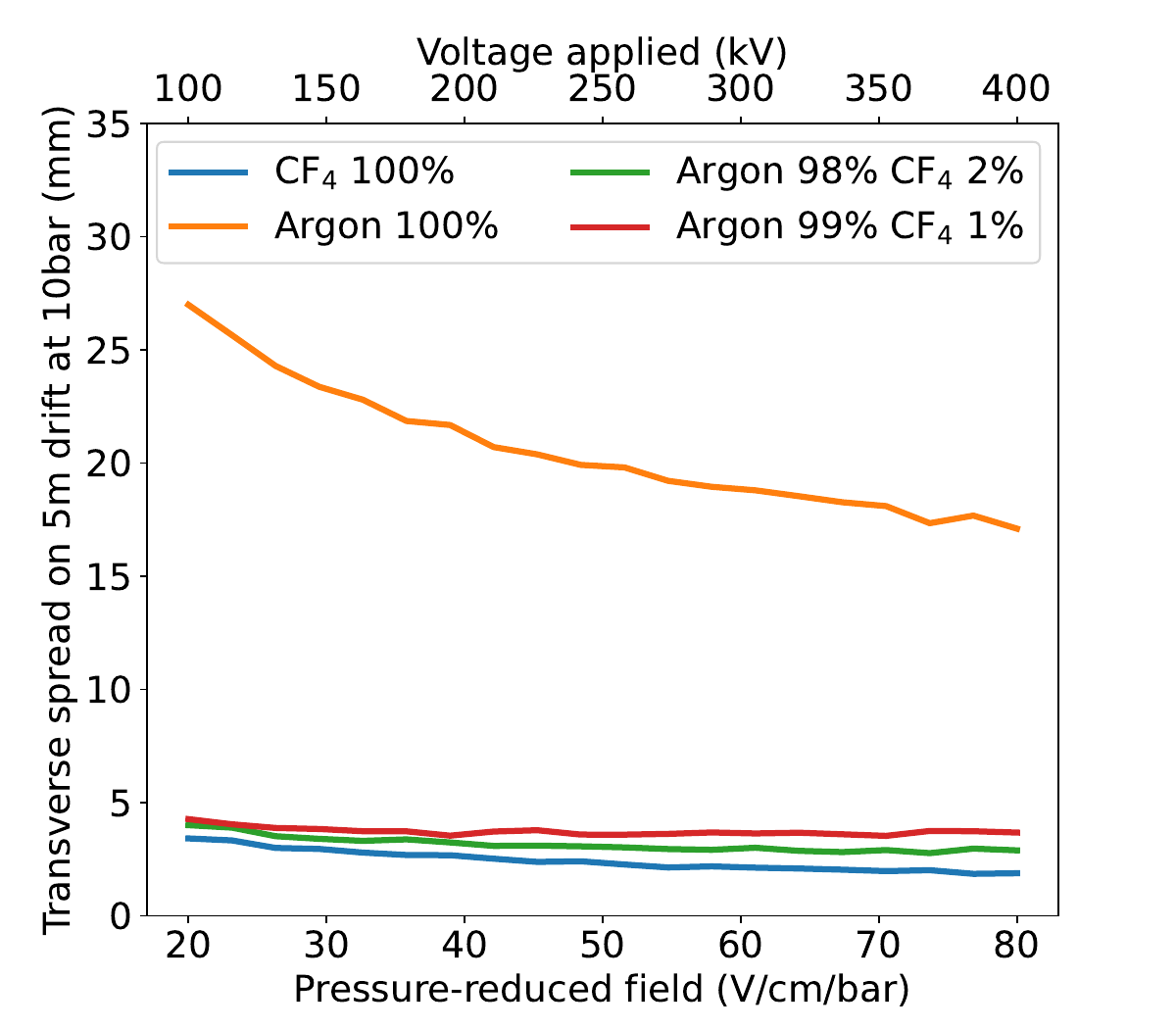}
\includegraphics[width=0.495\textwidth]{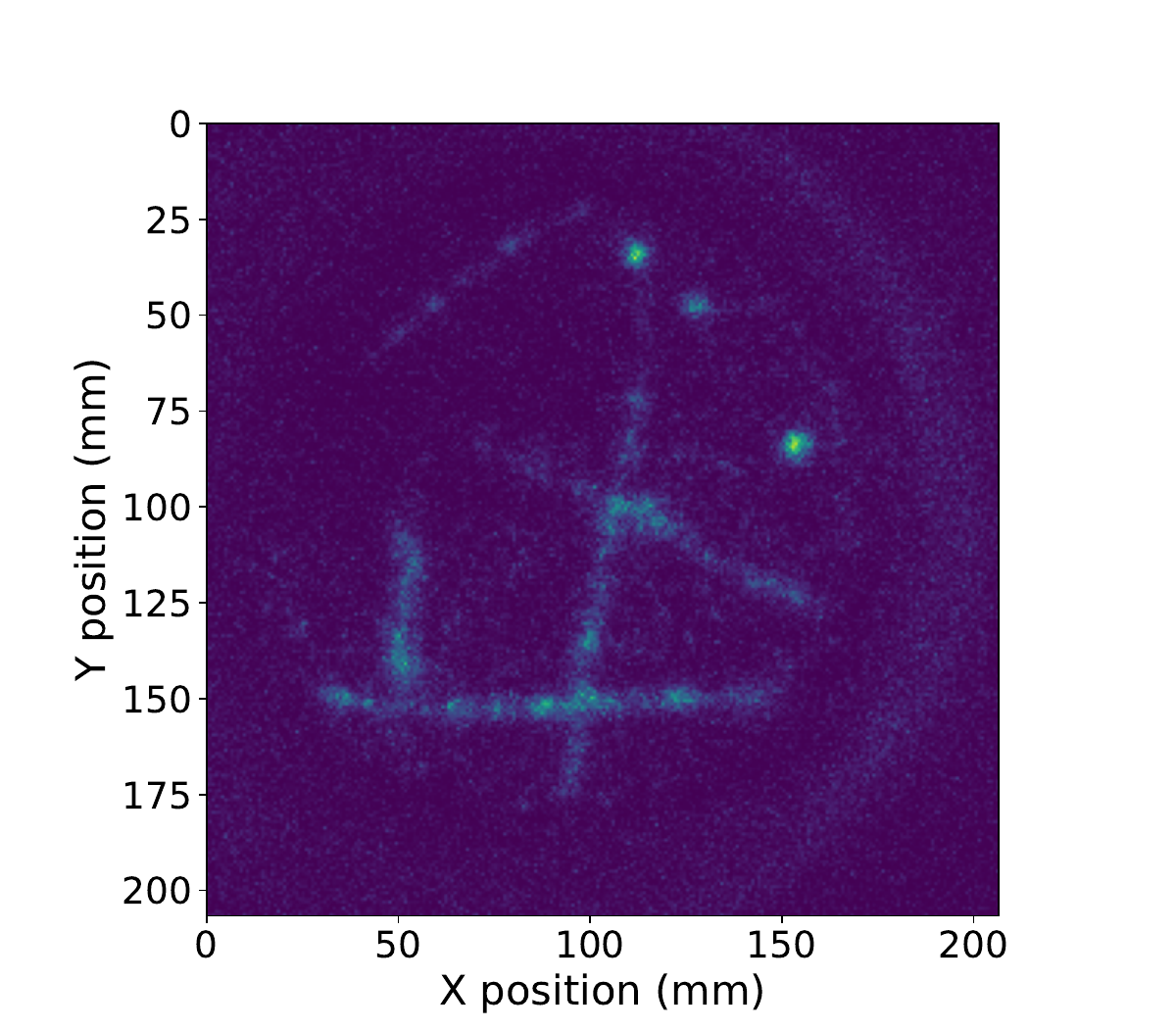}
\caption{Some fundamentals relevant to the present work. From top-left to bottom-right: (i) scintillation time profiles under 5.5~MeV $\alpha$ particles for Ar/CF$_4$ (99/1) at 9.5~bar (adapted from ref.~\cite{Amedo:2024skj}); (ii) emission spectrum of Ar/CF$_4$ in the visible region, per unit of energy deposited in the gas, obtained with an X-ray tube (adapted from \cite{Amedo:2023far}); (iii) transverse spread of the primary ionization after 5~m drift in different Ar and Ar/CF$_4$ mixtures, as obtained with Pyboltz \cite{AlAtoum:2019zpy}; (iv) cosmic ray tracks obtained in Ar/CF$_4$ (99/1) at 1~bar with a double-stack of glass-THGEMs \cite{Amedo:2023pmh}.}
\label{fig:CF4_Ref}
\end{figure}

Recently, high wavelength-shifting efficiency has been demonstrated for Ar/CF$_4$ mixtures \cite{Amedo:2023far, Amedo:2024skj}, resulting, for a mere 1\% volume fraction of CF$_4$, in scintillation strengths of about 1400~photons per MeV in the range between 400 and 700~nm (figure~\ref{fig:CF4_Ref}, top row). Such small additive concentrations are an important asset in order to avoid the contamination of the neutrino-Ar scattering data with spurious interactions from other nuclear species. Simulations performed with Pyboltz \cite{AlAtoum:2019zpy} support the fact that, at 400~V/cm, the strong electron cooling of CF$_4$ would reduce diffusion down to $\sigma_\mathrm{T}=3.6$~mm for a drift length of 5~m (fig.~\ref{fig:CF4_Ref}, bottom-left panel), a value even lower than what can be achieved with standard gas mixtures such as Ar/CH$_4$ at 90/10 (P10), for which $\sigma_\mathrm{T}=4.0$~mm is expected. According to Pyboltz, this performance is preserved in a wide range of drift fields (tens of V/cm/bar in a pressure-reduced representation). Although information on avalanche gain for low-quenched Ar/CF$_4$ mixtures is scarce, values in excess of $10^4$ have been reported in single-wire configuration \cite{Deptuch:2007ef}, and values around $3\times10^5$ in 3-GEM stacks down to CF$_4$ concentrations as low as 2\% at around atmospheric pressure \cite{DGD:2016}. The latter detector, instrumenting an area of $10\times10$~cm$^2$, provided high-quality $\alpha$ \cite{Brunbauer:2018xzt,Brunbauer:2018ebn} and X-ray  \cite{Brunbauer:2018nyz} images with reliable performance over the course of years, and is still in use at CERN's Gas Detectors Development laboratory. More recently, muon tracks have been imaged in Ar/CF$ _4$ (99/1) at 1~bar by resorting to a double glass-THGEM stack \cite{Amedo:2023pmh}, (fig.~\ref{fig:CF4_Ref}, bottom-right panel), with an optical gain around 10$^3$ obtained for each stage. The characteristics of CF$_4$ as a wavelength-shifting gas are convenient: it is relatively fast ---\thinspace most of the light is emitted with a time constant of about 15~ns \cite{Margato:2013gqa}, a feature preserved in the presence of Ar \cite{Amedo:2024skj}\thinspace--- and its emission spectrum lies largely in the near ultraviolet and visible regions, offering good prospects for detection. It is also highly immune to contaminants: at 1~bar, even a (certainly unrealistic) 3\% N$_2$ contamination would cause a tolerable $\times 2$ reduction of the CF$_4$ scintillation strength \cite{Margato:2012}.

Here we assess the potential of Ar/CF$_4$ mixtures for time-tagging in GArTPCs. The paper is structured as follows: section~\ref{sec:DetectorConcept} describes the detector concept and discusses which photosensor seems the fittest for our purposes given today's technological landscape; in section~\ref{sec:DetectorPerformance}, we present the Geant4 simulation we have developed to simulate light collection in the GArTPC and we discuss the main results for energy threshold, time and energy resolution for various event topologies; section~\ref{sec:technical} contains a technical discussion on
\begin{enumerate*}[label=(\roman*)]
\item the impact of the photosensor response, digitization scheme and electrical noise,
\item the performance with Winston cones, and
\item a simple proposal of an active cryostat aimed at achieving photosensor cooling in high-pressure closed systems, together with first experimental results validating the idea;
\end{enumerate*}
finally, in section~\ref{conclu}, we present a summary of the main results.

\section{Detector concept} 
\label{sec:DetectorConcept}
We will consider a cylindrical TPC with a diameter of 5~m and a length of 5~m, filled with Ar/CF$_4$ (99/1) at a pressure of 10~bar and a temperature of about 293~K, corresponding to a mass of argon of about 1.5~tons. Anode and cathode planes at the ends of the cylinder define a single-drift region. While the dimensions of such a chamber are comparable to those of the ALICE TPC \cite{Alme:2010ke, Lippmann:2014lay}, its active mass would exceed that of any existing or past gaseous TPC by at least one order of magnitude. The chosen geometry is inspired by the high-pressure TPC proposal of the DUNE experiment \cite{DUNE:2022yni}, that arguably sets the standard for future TPCs to be used as active targets in neutrino physics.

Regarding the readout of the primary ionization, we will assume hole-based amplification as in \emph{gas electron multipliers} (GEMs \cite{Sauli:1997qp,Sauli:2016eeu} or THGEMs \cite{Bressler:2023wrl}). The choice arises from the fact that, in the proposed concept, it is important to maximize the amount of primary scintillation while screening the one stemming from the avalanche process. Unlike open amplification structures (such as those based on wires \cite{Alme:2010ke}), modern hole-based micropattern gas detectors (MPGDs) can accomplish this to a large extent. For instance, quintuple \cite{Blatnik:2015bka} or dissimilar-pitch \cite{Lippmann:2014lay} GEM stacks are known to suppress feedback levels with minimal degradation in performance, while aluminium-electrode GEMs \cite{Chernyshova:2019} may be used to enhance the collection of primary scintillation from the main TPC volume. Even though other solutions might be developed along the way, we will focus in this work on demonstrated concepts. 

Regarding the readout of the primary scintillation, we will assume, for simplicity, that photon detection occurs behind a highly transparent cathode plane. Covering the field cage with photosensors is, in principle, possible \cite{nEXO:2018ylp}, despite additional technical difficulties: charging-up and electric field distortions, need of high voltage insulation and temperature stabilization, complexity of bringing services and reading out signals, and cost. For a tracking TPC in which most of the tracks would exit precisely through the field cage, there is no existing implementation that backs this idea. Therefore, we opt at this point for evaluating a more conventional solution where the field cage is lined with reflective materials (see, e.g., \cite{NEXT:2020pbx,NEXT:2022cmg} and references therein). 

There are, in principle, several large-area photosensors that could be suitable for the task proposed, including photomultiplier tubes (PMTs), microchannel plates (MCPs), avalanche photodiodes (APDs), CMOS or CCD cameras, and silicon photomultipliers (SiPMs). Examination of the main technical caveats seem to favour one of these options, as long as the discussion is restricted to demonstrated technological solutions:
\begin{itemize}
    \item PMTs: when it comes to large aperture, there is currently no commercial PMT above 1~inch that is rated for 10~bar \cite{Hamamatsu}. Moreover, ensuring magnetic-field compatibility inside the intended 0.5~T field would require dedicated R\&D. Lastly, the quantum efficiency (QE) of vacuum photocathodes in the red region of the spectrum is largely below what is achievable with silicon devices, and sits generally below 10\%.
    \item MCPs: compared to PMTs, microchannel plates have better immunity to magnetic fields, but they suffer too from the lack of availability of high pressure devices and low QE in the visible range, as PMTs do. For both PMTs and MCPs, detection of the UV-component of Ar/CF$_4$ might offer a better alternative in case of vacuum photocathodes, but remains to be studied.
    \item APDs: as shown in section~\ref{sec:DetectorPerformance}, in the present concept the scintillation will be generally down to single-photon levels per sensor, thus making this option non-viable.
    \item CMOS or CCD cameras: with a QE as high as 70\% or more in the visible spectrum and coupled to suitable optics, they can image large areas; however, they cannot be realistically used for the small signals expected in our case (hundreds of photoelectrons over the entire readout plane) due to simple solid-angle considerations.
    \item SiPMs: they can operate under high pressure and high magnetic fields, and their quantum efficiency can reach 30\% at 600~nm. Historically, SiPMs have suffered from high dark count rates (DCR). However, improvements over the last decade brought it down to the level of 50--100~$\mathrm{kHz}/\mathrm{mm}^2$ at room temperature \cite{NepomukOtte:2016ktf,HamamatsuMPPC:2023}, with ongoing efforts to bring it even further down \cite{ECFA:2021}. SiPMs are typically small, and hence not optimal for large-area coverage unless several channels can be combined (``ganged''). Their large gain facilitates this process, and signal-to-noise ratios better than 10 have been achieved for single-photon signals when, for instance, ganging together twenty-four $10\times10$~mm$^2$ SiPMs \cite{DIncecco:2017bau}.
\end{itemize}

In summary, an array of SiPMs tiling the cathode could realistically instrument the proposed TPC as long as the photosensors' DCR can be kept at tolerable levels. A common solution involves operation at low temperature, as the DCR is reduced by about an order of magnitude every 25~\celsius\ \cite{NepomukOtte:2016ktf,HamamatsuMPPC:2023}:
\begin{equation}
\mathrm{DCR} \simeq 100~\mathrm{kHz/mm}^2 \cdot 10^\frac{(T-25~\celsius)}{25~\celsius}.
\end{equation}
As will be shown in section~\ref{sec:DetectorPerformance}, temperatures around $-25~\celsius$ seem sufficient for achieving the target performance in our case. These moderately-low local temperatures are not unknown to particle physics instrumentation, even with the rest of the detector operating comfortably at room temperature (e.g., \cite{PANDA:2008rpr}). Ongoing studies point to the fact that two PMMA windows ---\thinspace one at the cathode and another one next to the photosensors\thinspace--- would suffice at providing thermal insulation between the active volume of the TPC and the SiPM plane, keeping the cooling power at reasonable levels (see section \ref{sec:technical}). For the evaluation of the detector performance, we consider as a prospective SiPM the 14160/14161 series by Hamamatsu, with a photosensor arrangement in square tiles of 35~$\times$~35~cm$^2$ at a pitch of 40~cm, leading to a maximum fill factor of $76.5\%$. The additional 5~cm provides space for the frames of the cryostat assembly, greatly reducing the thermal stress on the PMMA window. Conservatively, the reference simulations presented here have been done for half of the maximum fill-factor estimated this way (i.e., $38\%$, 7.5~m$^2$ photosensor area).

\section{Detector performance} 
\label{sec:DetectorPerformance}

\subsection{Simulation}
\begin{figure}
\centering
\includegraphics[width=0.5\textwidth]{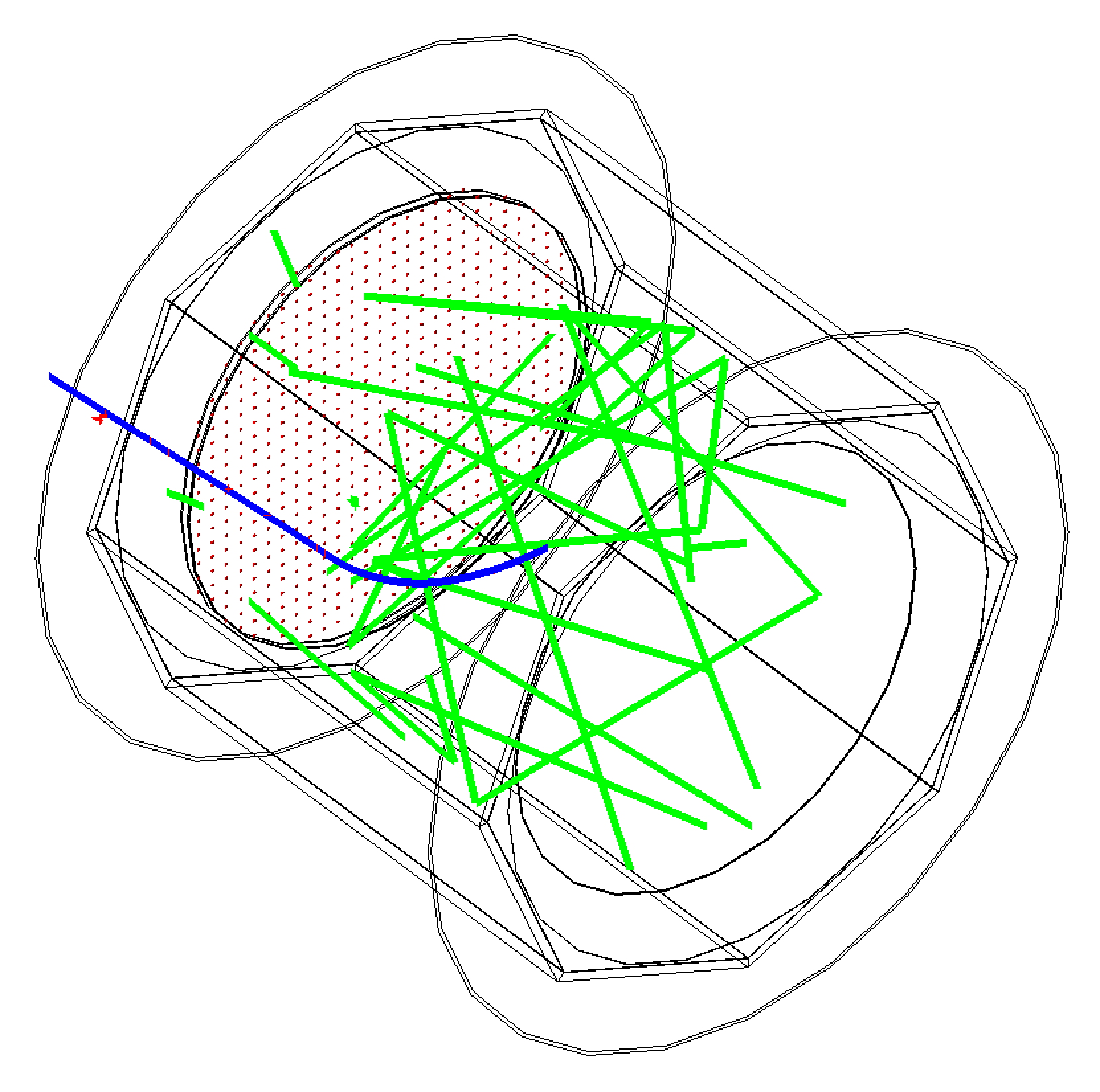}
\caption{Visualization of the detector geometry used in the Geant4-based simulation. The blue track represents the trajectory of a proton generated in the center of the TPC vessel, curved by the 0.5~T magnetic field present in the detector. The green lines represent optical photons produced by scintillation, which are reflected on the Teflon-coated inner surface of the field cage until they are detected by the SiPMs (depicted by the grid of red dots) or absorbed elsewhere. The octagonal structure represents an electromagnetic calorimeter (separated by a buffer region from the TPC) used to detect neutral particles escaping the TPC.}
\label{fig:simgeo}
\end{figure}

We evaluated the expected performance of the TPC concept described above by means of a Geant4 \cite{Allison:2016lfl, Allison:2006ve, GEANT4:2002zbu} end-to-end simulation that covers from the generation and transport of particles interacting in the detector, the production of the primary scintillation photons and, finally, their tracing (collection), amplification and conversion into electric signals. It implements a detailed optical model of the TPC geometry (see figure~\ref{fig:simgeo}), for which we will consider three main configurations (see figure~\ref{fig:scheme}, top row): (A) no reflectors; (B) a PTFE-lined field cage (with exposed field shapers, to minimize charging-up) and a fully-absorbing anode; (C) a PTFE-lined field cage and an anode based on aluminized GEMs. 

The primary scintillation spectrum of Ar/CF$_4$ mixtures \cite{Amedo:2023far} follows closely the one of pure CF$_4$ \cite{Morozov:2010}, with three main emissions: two in the UV range (230~nm and 280~nm), attributed to CF$_4^{+,*}$ transitions, and one in the visible range (630~nm), attributed to a CF$_3^*$ transition. Given the better prospects of the visible emission for the detection of low-energy, highly-ionizing radiation \cite{Amedo:2023far}, only the latter component will be considered in the following. The scintillation time profile for Ar/CF$_4$ gas (at 99/1, 10~bar) was taken from the same work (fig.~\ref{fig:CF4_Ref}, top-left panel). The effective reflectivity of the PTFE-covered drift wall, which assumes aluminium field spacers as inserts in a 20:1 ratio (95$\%$ fill factor), is 94.5\%. This effective value results from the weighted reflectivity of the PTFE, $\sim95\%$ \cite{Janecek:2012,NEXT:2020pbx}, and of the aluminium, $\sim85\%$ \cite{Lindseth:1999ole}, in the wavelength range of 550--700~nm. The effective reflectivity of the GEM-based anode depends on the ratio of the hole diameter to the hole pitch. Taking the large-pitch GEMs from the ALICE TPC \cite{Lippmann:2014lay} as a reference (70~\textmu{m} hole diameter, 280~\textmu{m} pitch), the effective GEM reflectivity in case of aluminium electrodes may be up to $\sim80.8\%$. For more conventional GEMs (70~\textmu{m} hole diameter, 140~\textmu{m} pitch), the resulting reflectivity is $\sim68.3\%$.

\subsection{Light collection efficiency}
\begin{figure}
\centering
\includegraphics[width=\textwidth]{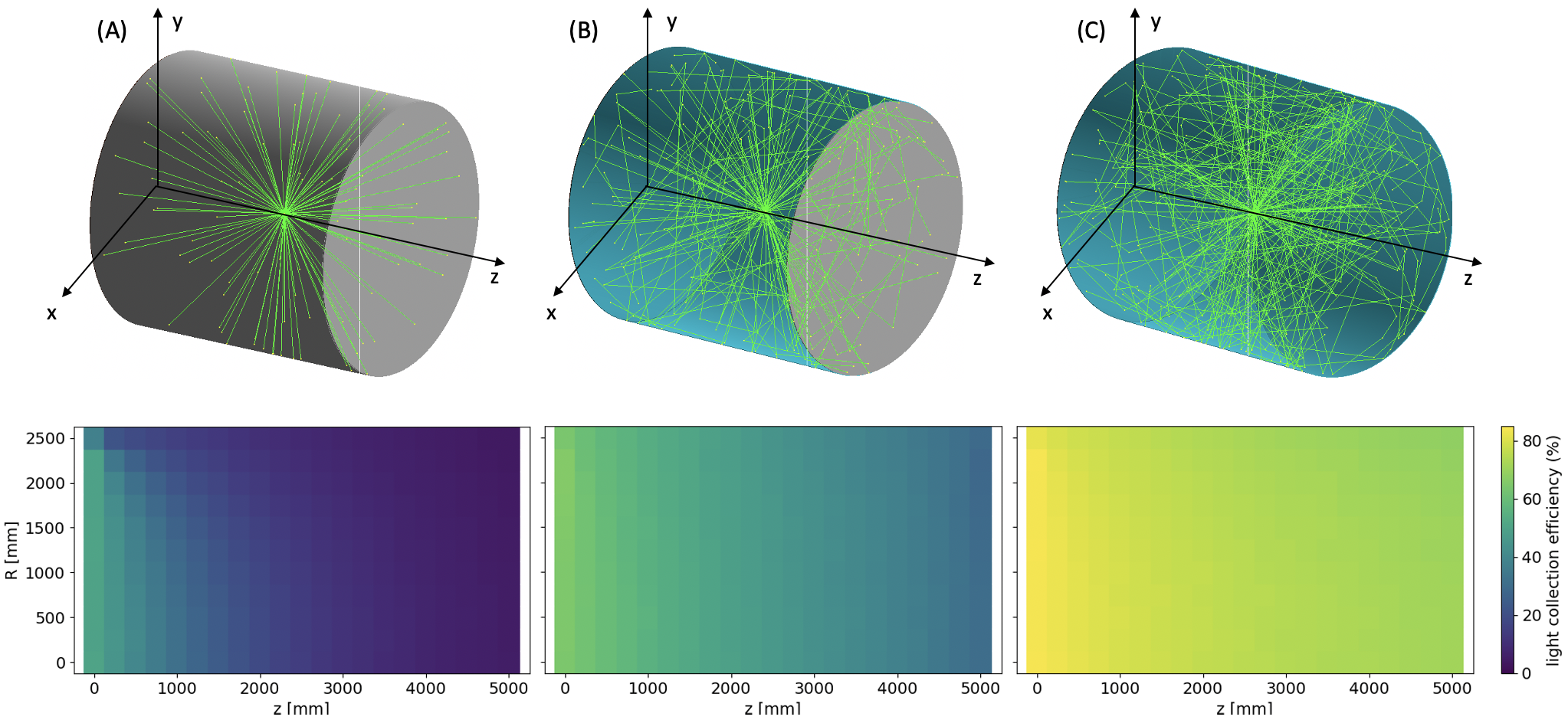}
\caption{Results from photon tracing for the three TPC configurations discussed, where reflective surfaces are represented in blue and non-reflective ones in grey: (A) TPC without reflective lining; (B) TPC with a PTFE reflector covering the interior of the field cage; and (C) like case B, but with aluminized GEMs instrumenting the anode. The plots on the second row show the light collection efficiency as a function of the radial and axial positions of a point-source for each of the configurations. The axial coordinate $z$ is taken to be zero at the photosensor plane.} \label{fig:scheme}
\end{figure}

\begin{figure}
\centering
\includegraphics[width=0.60\textwidth]{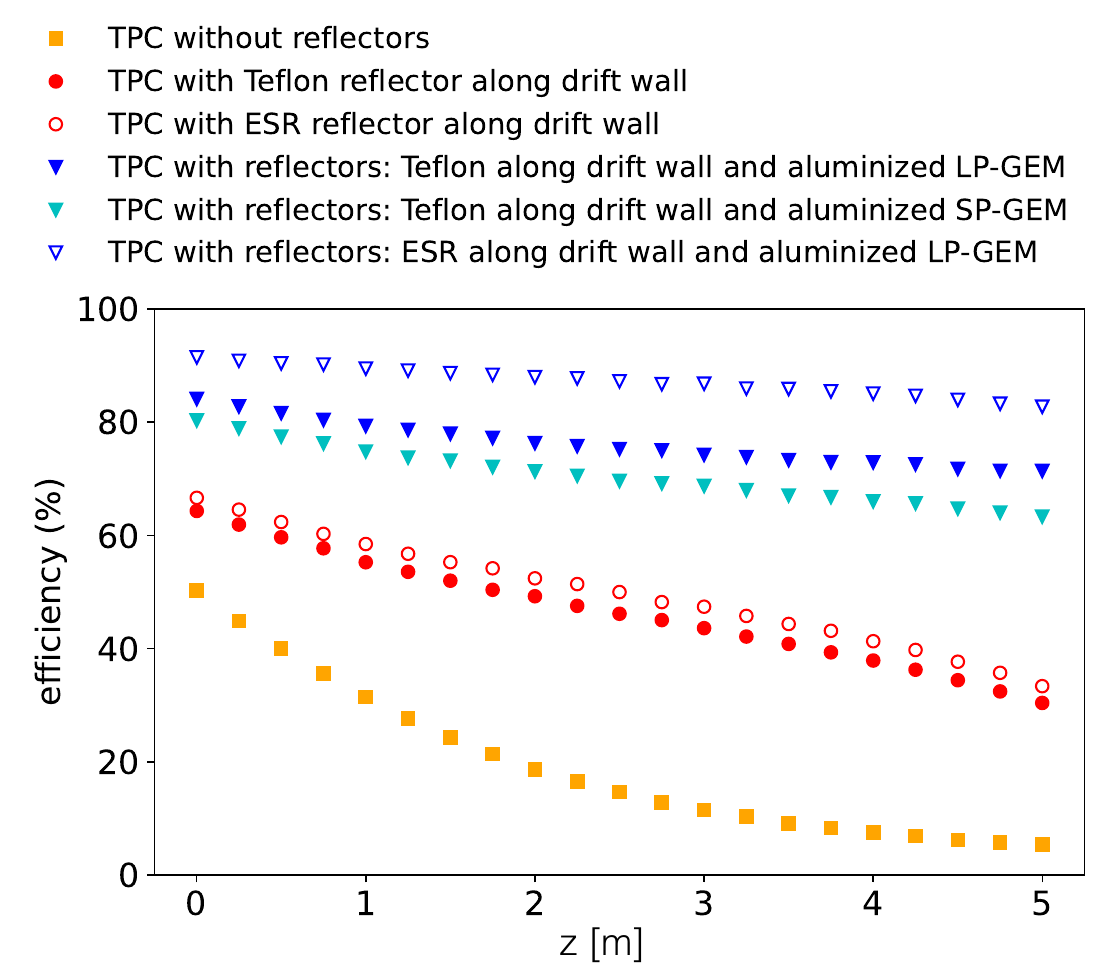}
\caption{Light collection efficiency ($\eta$) for the three main TPC configurations discussed in text: (A) TPC without reflective lining (orange squares); (B) TPC with Teflon (or ESR reflector) in the field cage (red circles); (C) like case B, but with aluminized GEMs instrumenting the anode (blue triangles). The axial coordinate $z$ is taken to be zero at the photosensor plane.}
\label{fig:collection_efficiency}
\end{figure}

Using our Geant4 simulation, we computed the light collection efficiency ($\eta$) of the detector, defined as the number of photons reaching the photosensors (i.e., excluding their photon detection efficiency) divided by the number of generated ones. It is shown as a function of cylindrical coordinates (radial and axial position) in figure~\ref{fig:scheme} for the three TPC configurations listed above. To simplify the discussion, figure~\ref{fig:collection_efficiency} shows $\eta$ for different positions of a point source placed along the central axis of the TPC. The collection efficiency decreases fast towards the anode (higher $z$) in the absence of reflectors (square data points), flattening when reflectors are included in the field cage (red circles) and at the anode (blue triangles). Differences in $\eta$ obtained by replacing Teflon (a brand name for PTFE) with specular reflector film (ESR) ---\thinspace with average reflectance around 98\% in our range of interest; see, e.g., refs.~\cite{Janecek:2012,Okumura:2017qvd}\thinspace--- or using either large-pitch (LP) aluminized GEMs (dark blue triangles) or conventional standard-pitch (SP) ones (light blue triangles) are all well within 10 percentage points. Hence, without resorting to special techniques, light collection values in the range 70--90\% may be reached over the entire chamber, showcasing the convenience of using visible light for detection.

The average number of photoelectrons ($N_\mathrm{pe}$) expected from a physics signal is given, approximately, by the following expression: 
\begin{equation}
N_\mathrm{pe} \simeq Y_\mathrm{sc} \cdot \eta \cdot T_\mathrm{PMMA} \cdot F_\mathrm{SiPM} \cdot \varepsilon_\mathrm{SiPM}, \label{eq:Npe}
\end{equation}
where $Y_\mathrm{sc} = 1400~\mathrm{photons/MeV}$ (as discussed in section~\ref{sec:Introduction}) is the assumed scintillation yield in the visible range for Ar/CF$_4$ (99/1) at 10~bar, that is, the average number of photons generated by a primary particle per unit of deposited energy; $\eta$ is the light collection efficiency; $T_\mathrm{PMMA}\simeq0.9\cdot0.9=0.81$ is the optical transmission of the two PMMA windows in front of the photosensors; $F_\mathrm{SiPM} = 0.38$ is the fill factor of the SiPM array; and $\varepsilon_\mathrm{SiPM}=0.255$ is the average photon detection efficiency (PDE) of the SiPMs in response to the visible component of Ar/CF$_4$. For a point-like event (e.g., a proton track of a few MeV) in the center of a Teflon-lined TPC (configuration B), $\eta=47\%$ is obtained (red solid circles) and, hence:
\begin{equation}
N_\mathrm{pe}\simeq 52~\mathrm{photoelectrons/MeV},
\label{NpheMeV}
\end{equation}
ranging between 38 and 71 for events in the vicinity of anode and cathode, respectively.

\subsection{Pulse shape and pulse reconstruction}
\begin{figure}
\centering
\includegraphics[width=\textwidth]{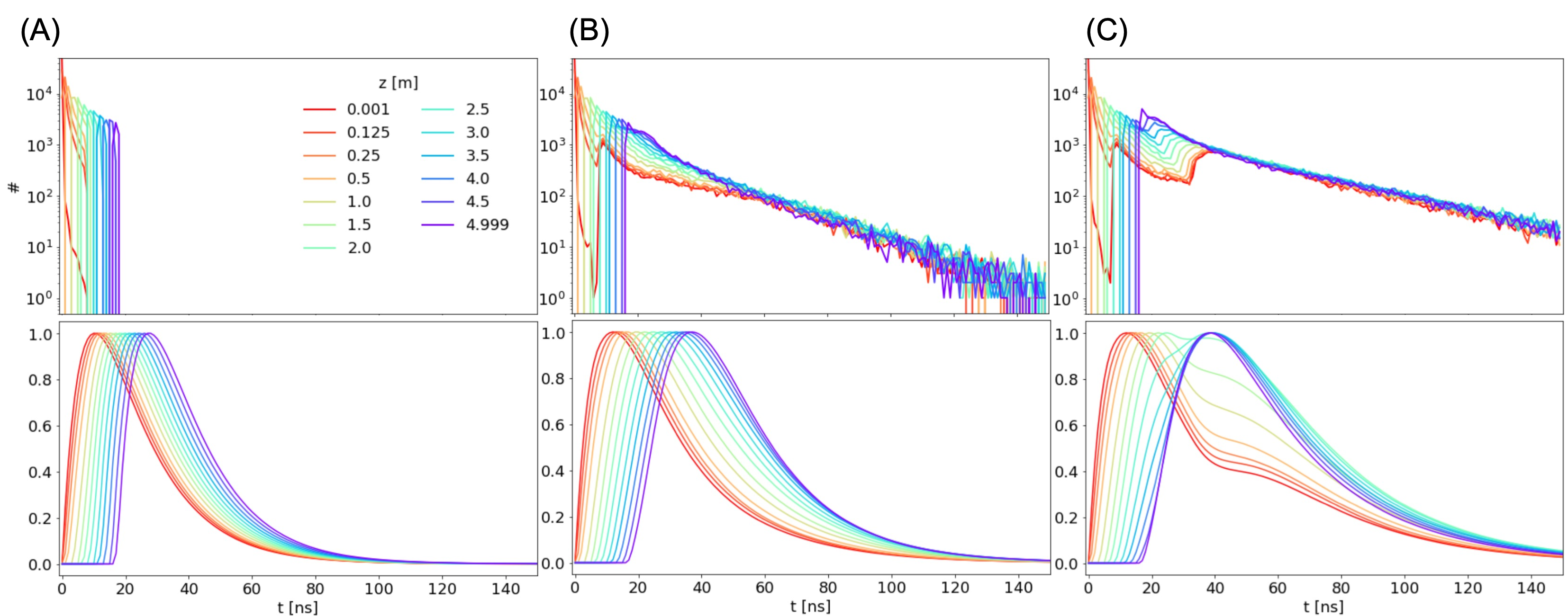}
\caption{Top: Geant4 simulation of the photon arrival times for protons of 5~MeV in the three reference configurations discussed in the main text. The contribution from the photons reflected from the field cage and anode appear distinctly separated in case C up to about mid-chamber. Bottom: same simulation but including the gas scintillation response. Pulse shapes in a fully reflective TPC (last column) show a strong dependence with the position of the interaction, unlike the others.}
\label{fig:timesignal}
\end{figure}

The time response of the TPC results from the time profile of the gas scintillation (fig.~\ref{fig:CF4_Ref}, top-left panel) coupled to the distribution of arrival times at the photosensor, including all reflections. Geant4-simulated time profiles for point-like energy deposits are shown in figure~\ref{fig:timesignal} for the configuration without reflectors (A), reflective field cage (B) and reflective field cage and anode (C), and for different $z$ positions of the light source within the chamber. The top-row panels show the time profiles from photon-tracing, while the bottom row shows the full TPC response including the gas scintillation time profile (arbitrarily normalized to 1).

In the following, time and energy resolution will be assessed through an event-by-event analysis, fitting the simulated pulses to signal templates (obtained from high-statistics simulations, as depicted in fig.~\ref{fig:timesignal}), and leaving the pulse time and amplitude free. It is shown later, in section \ref{sec:Electronics}, that a sampling of 4--5~ns (about twice the Nyquist sampling rate in present conditions) preserves the pulse information, and it has been chosen hereafter. To avoid discretization artifacts, the time sampling is set to start asynchronously, event by event.

\begin{figure}
\centering
\includegraphics[width=0.75\textwidth]{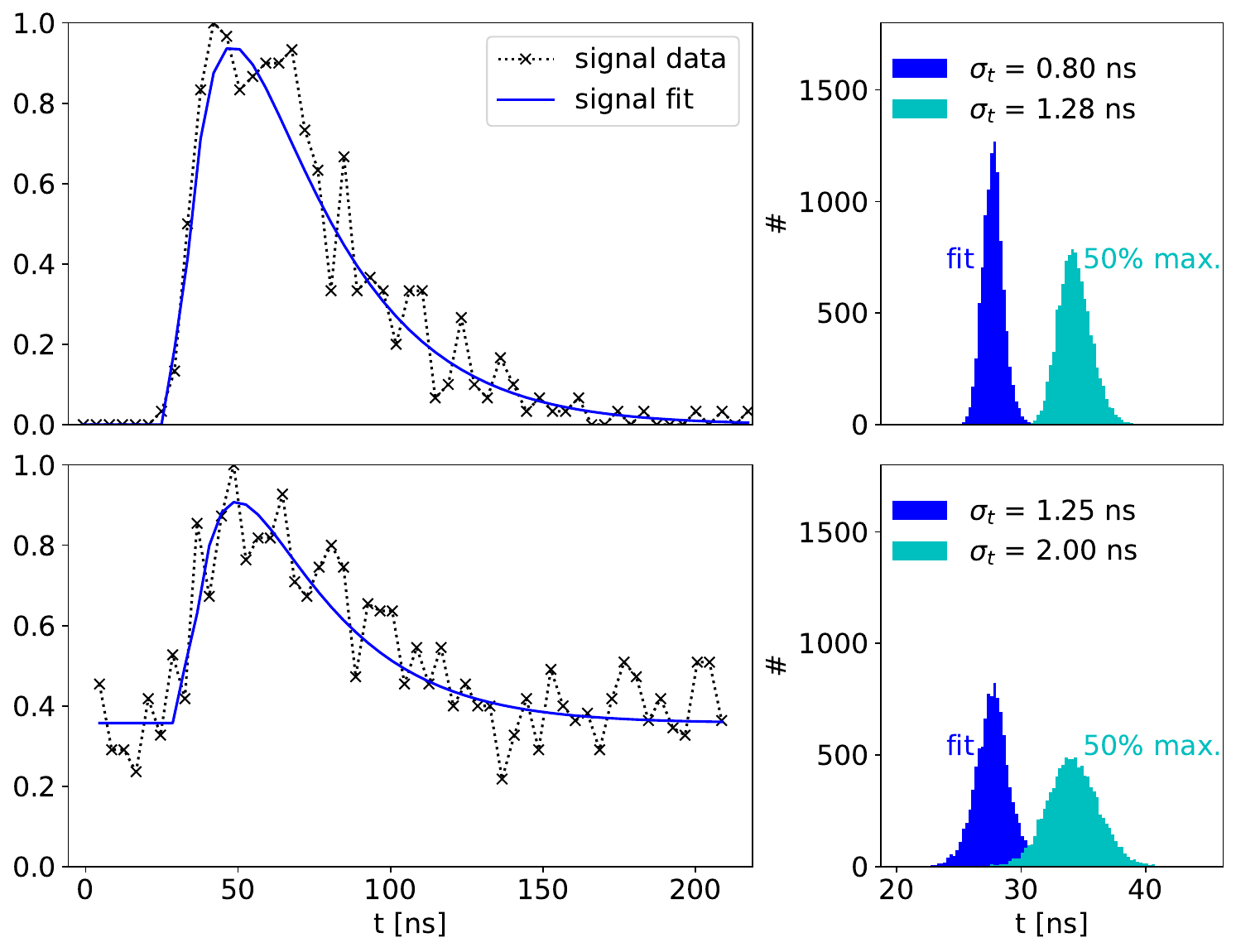}
\caption{Left: example of pulse reconstruction at 4-ns steps for a point-like signal of 5~MeV with no DCR from the SiPMs (top) and with a DCR corresponding to operation at $T~=~-25^\circ$C (bottom). Right: histogram showing the time resolution obtained from the reconstruction of 10\,000 pulses when using a fit (dark blue) and from a fixed fraction (50\%) of the peak height (cyan). Pulses were arbitrarily normalized to one.}
\label{fig:pulsereconstruction}
\end{figure}

Examples from the fitting procedure for 5~MeV deposits under different dark count levels are shown in figure~\ref{fig:pulsereconstruction} (left), which also provides (right panel) the distribution of reconstructed pulse-times obtained from the fit. It should be noted that, in the absence of experimental data, it is difficult (and probably futile) to evaluate the extent to which the shape of the pulse must be known prior to performing the peak-fitting routine. However, as suggested by fig.~\ref{fig:timesignal}, and confirmed later for extended tracks, the TPC concept described here leads to very similar scintillation pulses for any particle type and position within the chamber. As a worse-case scenario, fig.~\ref{fig:pulsereconstruction} also shows the distribution of reconstructed times when considering a shape-independent estimate (in this case, the time at 50~\% of the signal maximum). For reference, in configurations A and B, a simpler template based on a bi-exponential signal convoluted with a Gaussian response of width $\sigma$ and zero mean ($\mathcal{G}(t,\sigma))$ was employed too: 
\begin{equation}
f(t) = A \left(e^{-(t-t_0)/\tau_1} - f_{21} \cdot e^{-(t-t_0)/\tau_2} \right) * \mathcal{G}(t,\sigma). \label{Eqpulse}
\end{equation}
This leads to very similar results compared to the exact template.

\subsection{Detection threshold}
\begin{figure}
\centering
\includegraphics[width=0.55\textwidth]{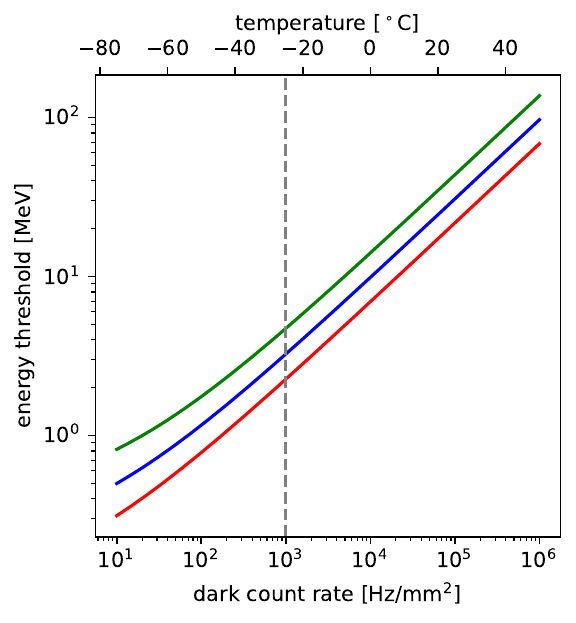}
\caption{Minimum event energy required to produce a detectable primary-scintillation signal (5$\sigma$ significance), for a Teflon-lined TPC filled with Ar/CF$_4$ (99/1), as a function of the SiPM DCR (bottom axis) and the corresponding operating temperature of the SiPM plane (top axis). For reference, a point-like event at about mid-chamber has been considered. The three lines correspond, respectively, to 76\% (red), 38\% (blue) and 19\% (green) sensor coverage of the anode plane.}
\label{fig:energy_threshold}
\end{figure}

The minimum particle energy needed to enable detection of the primary scintillation signal can be estimated by defining the 5$\sigma$ sensitivity of the SiPM array as the average upper limit (computed here using the Feldman-Cousins frequentist prescription), that would be obtained by an ensemble of measurements with the expected background (the photosensor's DCR) and no true signal. Figure~\ref{fig:energy_threshold} shows the energy threshold as a function of the DCR, for an event occurring at around mid-chamber in configuration B (Teflon-lined TPC), and with different photosensor coverages.

\begin{figure}
\centering
\includegraphics[width=0.80\textwidth]{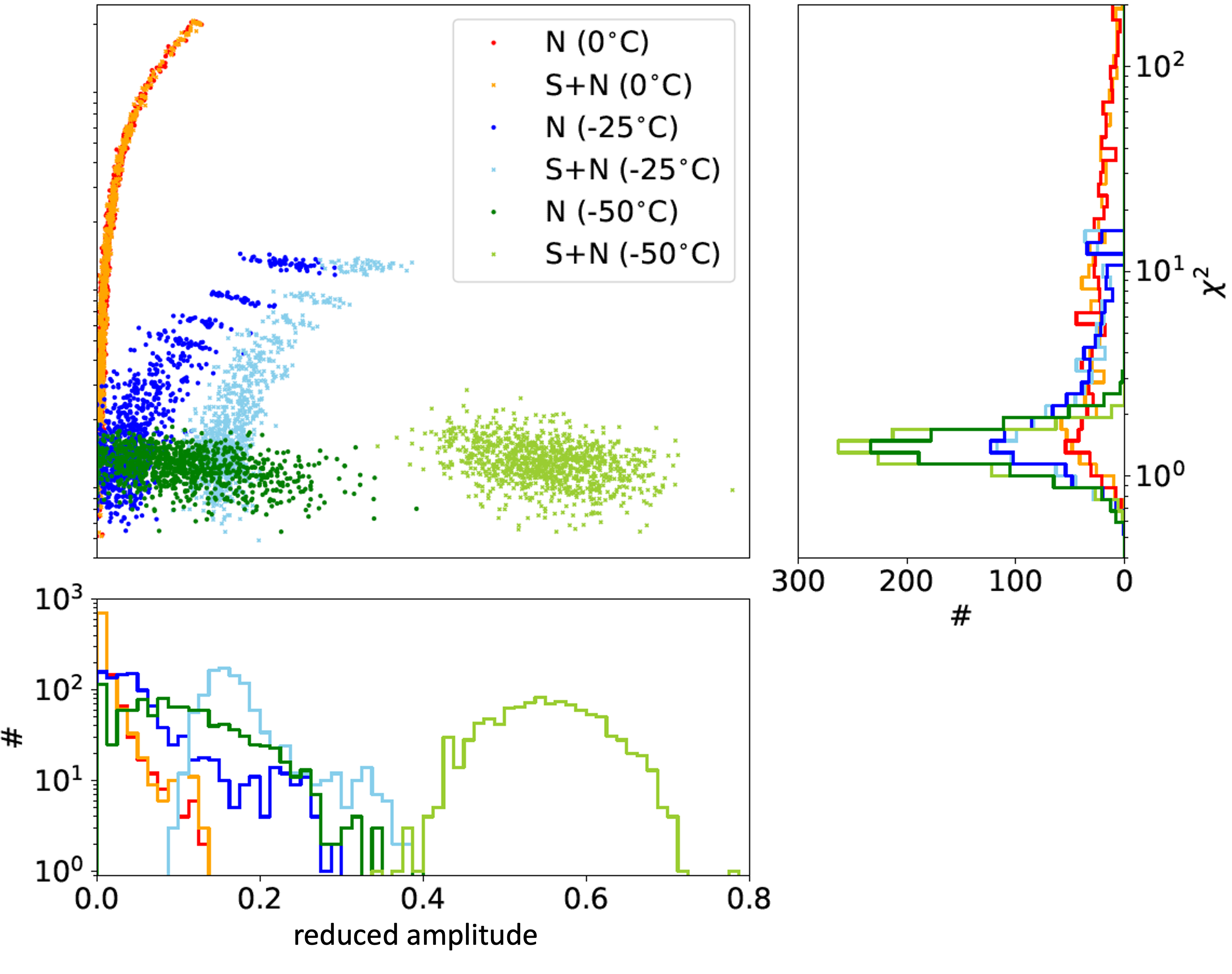}
\caption{$\chi^2$ vs amplitude for 5 MeV energy deposits at around mid-chamber for a Teflon-lined TPC (configuration B). Different temperatures of the photosensor plane are considered in the reconstruction of \emph{noise} ($N$) and \emph{signal-plus-noise} ($S+N$) events. For clarity of representation, the $x$ axis represents the signal amplitude divided by the sum of amplitude and baseline.}
\label{fig:chi2_amplitude}
\end{figure}

Another way to analyze the situation, without any prior information on the arrival time of the pulse, is to consider how the reconstruction described earlier performs both on a baseline given by dark rate alone and on an actual pulse superimposed on such a baseline (chosen to be 200~ns wide). The obtained $\chi^2$ vs amplitude plots allow the determination of when a signal can be confidently identified (figure~\ref{fig:chi2_amplitude}): for 5 MeV energy deposits at the center of a Teflon-lined TPC and a temperature of around $-25$~\celsius, both signal and baseline events can be well-separated (dark and light-blue dots), becoming indistinguishable at 0~\celsius\ (red and yellow dots). Going down to $-50$~\celsius, the two distributions become perfectly separated, even when considering the amplitude alone.

\subsection{Time and energy resolution for point-like tracks}
\begin{figure}
\centering
\includegraphics[scale=0.45]{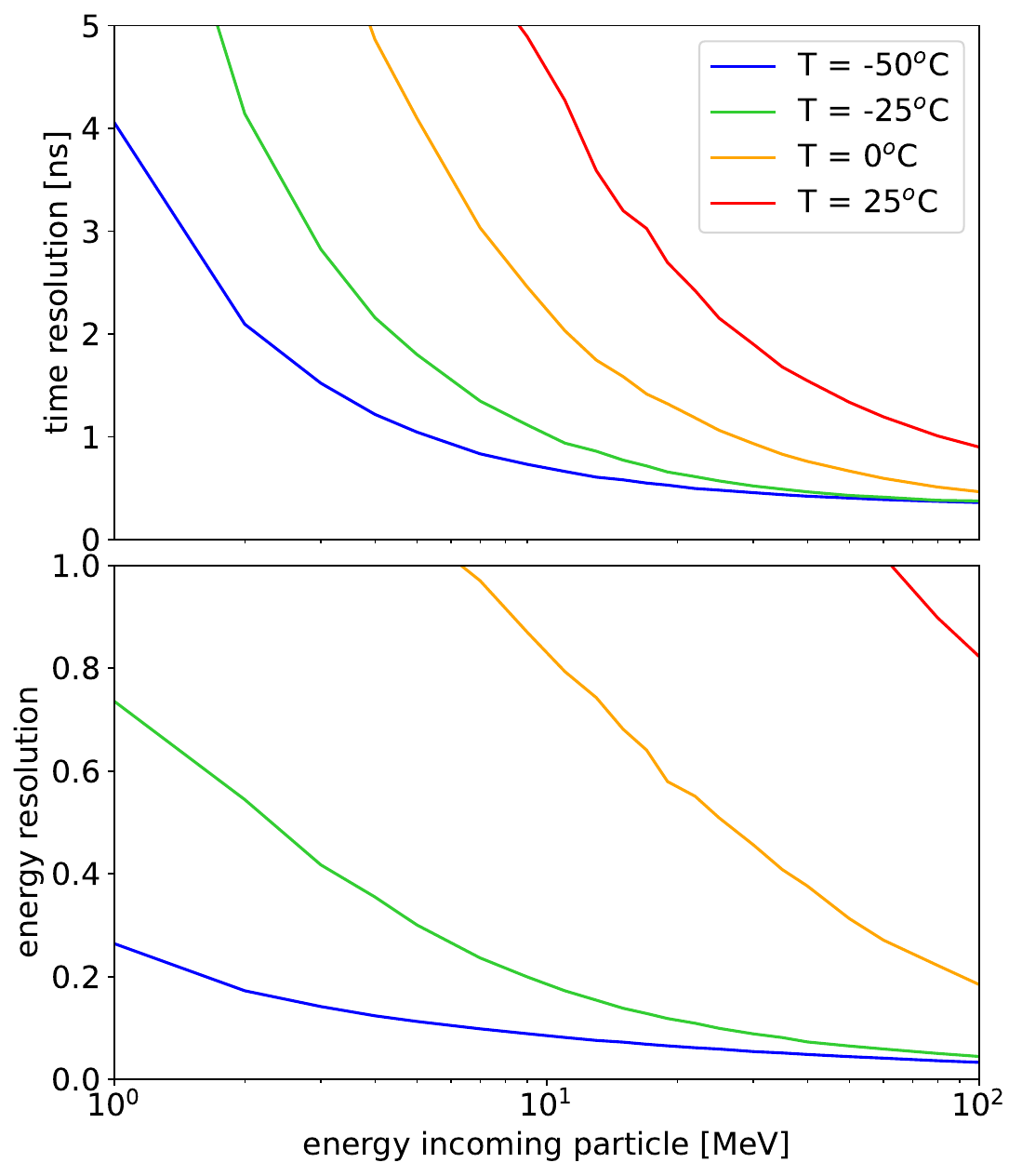}
\caption{Accuracy of reconstruction of the primary scintillation pulse in Ar/CF$_4$ gas (99/1) at 10~bar, in the 5 m-diameter, 5 m-length TPC considered in this work. Point-like deposits at mid-chamber and a Teflon-lined field cage have been assumed (configuration B). Results are presented as a function of energy and for various operating temperatures of the SiPM plane. Top: time resolution ($\sigma$). Bottom: energy resolution ($\sigma$/mean).}
\label{fig:t0_vs_E}
\end{figure}

The quality of the pulse-reconstruction procedure can be judged from the accuracy with which the event energy and time can be retrieved. The top panel of figure~\ref{fig:t0_vs_E} shows the spread ($\sigma$) of the time distribution, while the bottom panel shows the relative spread ($\sigma$/mean) of the amplitudes obtained from the pulse fit as a function of the event energy. In both cases, a point-like event at mid-chamber is considered in a Teflon-lined TPC (B). In the background-dominated region, the performance improves as $\sim 1/\mathrm{energy}$ and tends asymptotically to $\sim 1/\sqrt{\mathrm{energy}}$, the point at which all curves merge.

Energy resolution is a relevant figure in a high multiplicity environment, given that energy information is one of the criteria for matching scintillation and ionization signals in order to assign the time of the interaction to an event (the other criterion being the spatial correlation between the ionization trails and the photon positions). The energy resolution from the ionization signal plays little role, as it is expected to be much better than in the scintillation channel. It improves approximately with the square root of the number of detected carriers, which is more than one order of magnitude higher in the ionization channel than in the scintillation one, in the gas conditions discussed. For reference, the average energy lost by a high-energy electron or muon resulting from a charged-current interaction in the gas, or from background interactions in the rock and upstream materials, is in the range of 9--11 MeV (assuming a 5 m track in Ar/CF$_4$ at 10 bar). The energy resolution is 18\% at $-25$~\celsius and 8\% at $-50$~\celsius for 10 MeV. While this makes it challenging to associate and time-tag isolated leptons without external tracking, interactions resulting in hadron tracks with energy above 15 MeV will be easily recognizable at 2$\sigma$ over the muon/electron field. The ability to perform pulse reconstruction in a high multiplicity environment is discussed in section \ref{sec:inspill}.

\begin{figure}
\centering
\includegraphics[width=\textwidth]{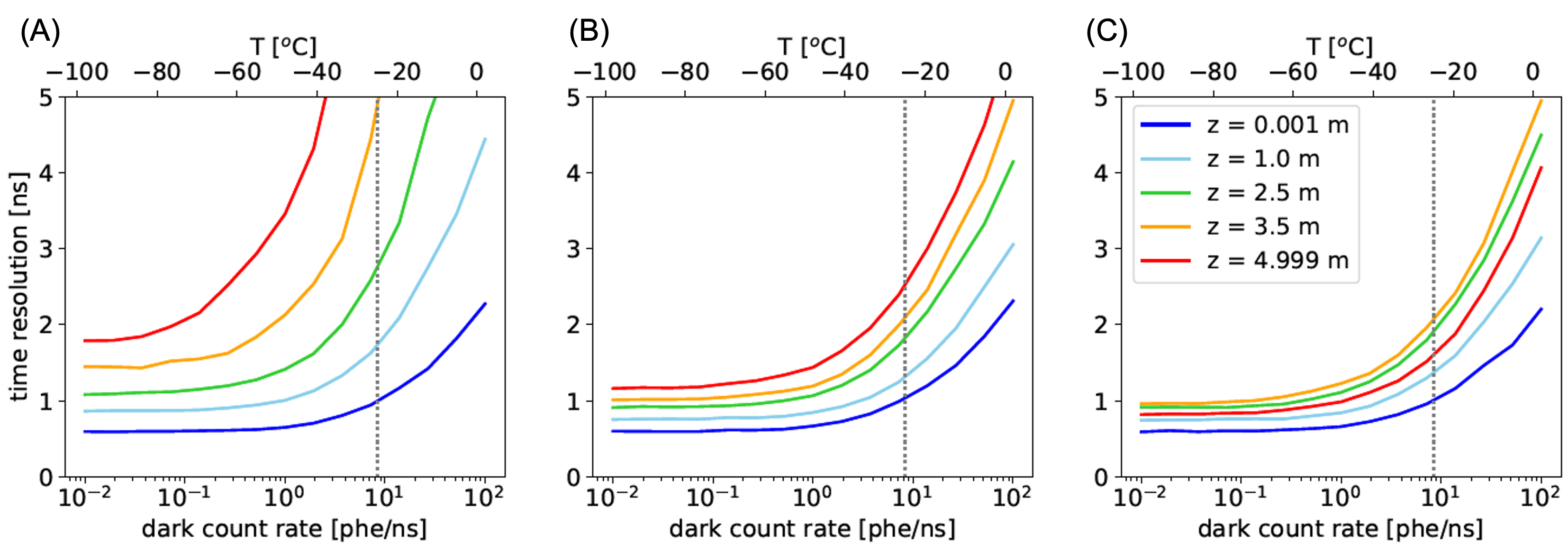}
\caption{Time resolution for the reconstruction of the scintillation pulse in Ar/CF$_4$ gas (99/1) at 10~bar, in the 5~m-diameter/5~m-length TPC considered in this work. It is represented for 5~MeV deposits as a function of DCR (lower axis) and temperature of the SiPM plane (upper axis), considering different distances $z$ to the photosensor plane and for three TPC configurations: (A) TPC with no reflectors; (B) with Teflon reflector along the drift wall; (C) with Teflon reflector along the drift wall and aluminized GEMs.}
\label{fig:t0_vs_dcr}
\end{figure}

When the matching between ionization and scintillation signals can be accomplished for an interaction inside the TPC, the time retrieved from the latter enables different capabilities, depending on the achievable resolution: (i) performing spill-association, (ii) $z$-determination ($\sim$ cm/${\mu}$s -scale in gas) and (iii) vertex assignment of neutral events (expectedly, the shortest time scale of practical use). In particular, the study in \cite{DUNE:2021tad} indicates that neutrons from neutral-current interactions can be well assigned to the interaction vertex if their time-of-flight can be measured to ns accuracy. Figure~\ref{fig:t0_vs_E} (top panel) indicates that this time resolution value may be reached for $\sim{10}~$MeV energy deposits at mid-chamber, if the SiPM plane is cooled down to $-25$~\celsius ~(0.75~ns at $-50$~\celsius). Focusing on this observable, fig.~\ref{fig:t0_vs_dcr} presents the time resolution as a function of the DCR (bottom axis), and the corresponding operating temperature of the SiPMs (top axis), for the three configurations discussed. The study has been done for point-like deposits of 5~MeV (a bit above the energy threshold estimated in previous sections). The dashed vertical line indicates the proposed operating temperature ($-25$~\celsius). 

It can be noted that, from the point of view of the time resolution, the addition of a reflector at the anode is of little advantage as the anode-reflected light contributes little to timing except for events close to the anode (red line). When using reflectors, time resolution is situated in the range 1--2.5~ns at $-25$~\celsius, whereas obtaining a comparable performance in the absence of them would require operation close to $T=-65$~\celsius. An improvement of about a factor of 2 can be obtained for operation at $T=-50$~\celsius\ in configurations B and C (centre and right panels in fig.~\ref{fig:t0_vs_dcr}), a performance that cannot be matched in the absence of reflectors no matter the operating temperature (left plot in fig.~\ref{fig:t0_vs_dcr}).

\subsection{Response to extended tracks}
\begin{figure}
\centering
\includegraphics[width=\textwidth]{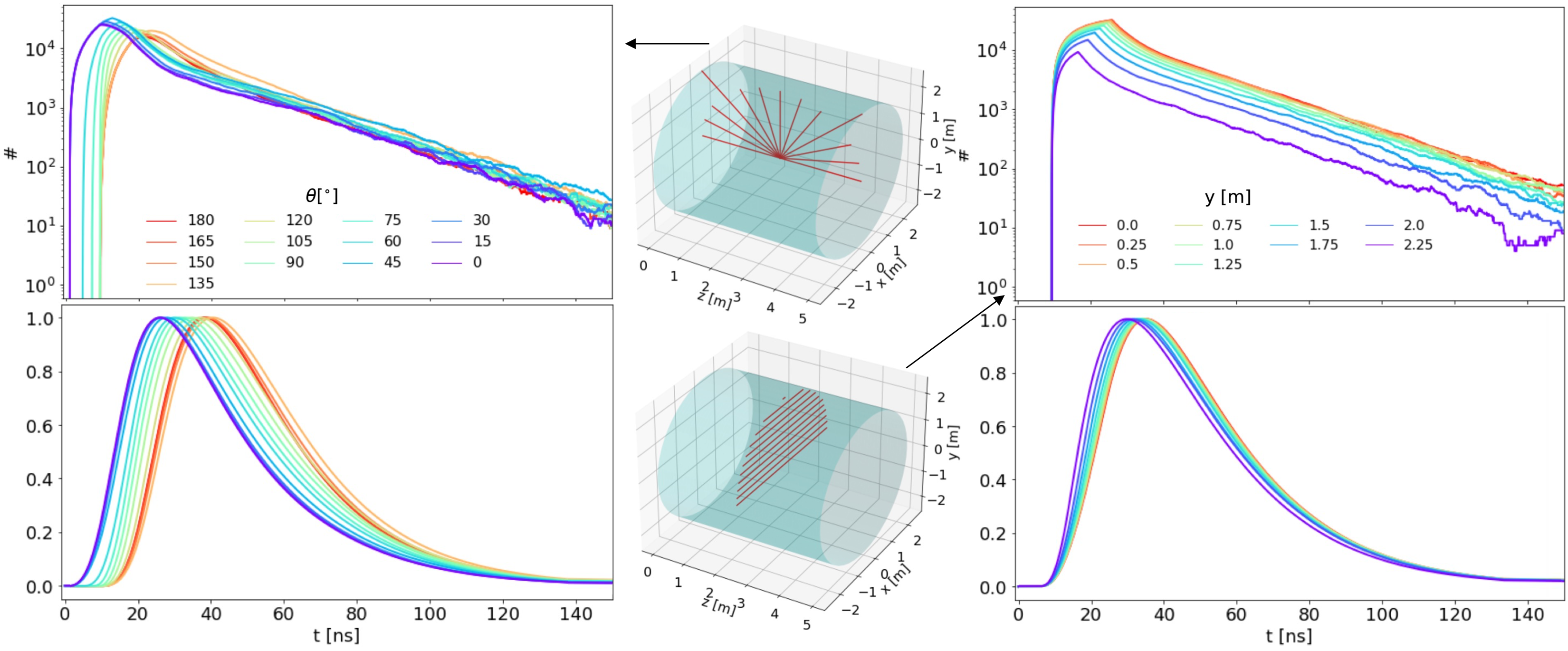}
\caption{Top: Geant4 simulation of the photon arrival times for inner muons (left) and external muons (right) in configuration B (Teflon-lined field cage). Bottom: Same simulation including the gas scintillation response (linear scale). The track topologies considered are sketched in the central column.}
\label{fig:timesignals_muons}
\end{figure}

Background muons from neutrino interactions in the rock upstream of the TPC can be expected, entering it at different heights ($y$) and values of the drift coordinate ($z$). On the other hand, leptons produced inside the chamber from charged-current interactions will be emitted at vastly different angles and interaction points. We choose such \emph{external} and \emph{internal} muons as our study case. Illustratively, figure~\ref{fig:timesignals_muons} shows the simulated time distributions before (top panel) and after (bottom) including the time profile of the Ar/CF$_4$ scintillation. Configuration B (Teflon-lined field cage) was chosen again for our investigations. For internal muons, tracks have been generated at mid-chamber in varying angles (left), whereas for external muons different track heights are considered, parallel to the cathode plane and at the middle of the TPC (right). Even as the Geant4 pulses are widened due to the time elapsed by the muon crossing the chamber, the scintillation profile is still the dominant effect in the final pulse shapes, that are hence very similar to the ones obtained for point-like particles in figure~\ref{fig:timesignal}. Similar to the case of 5 MeV point-like deposits, and due to the comparable energy loss (1.8--2.2~MeV/m for Ar at 10~bar), the time resolution values for muons with the geometries shown in fig.~\ref{fig:timesignals_muons} are also in the 1--2~ns range at $-25$~\celsius.

\subsection{In-spill reconstruction for high event multiplicities}
\label{sec:inspill}
The ability to reconstruct multiple events per beam spill depends on the width of the primary scintillation pulse, whose largest contribution comes from the scintillation time profile of Ar/CF$_4$, about 25~ns FWHM. To evaluate the reconstruction capabilities at high multiplicities, 10 to 100 muons with an energy of 1~GeV have been simulated, within a time window of 10~\textmu{s}. They are assumed to cross the detector from the upstream direction, within a uniform irradiation field (i.e., they are \emph{external} muons, according to the convention introduced earlier).

\begin{figure}
\centering
\includegraphics[width=\textwidth, trim={4cm 0 4cm 0},clip]{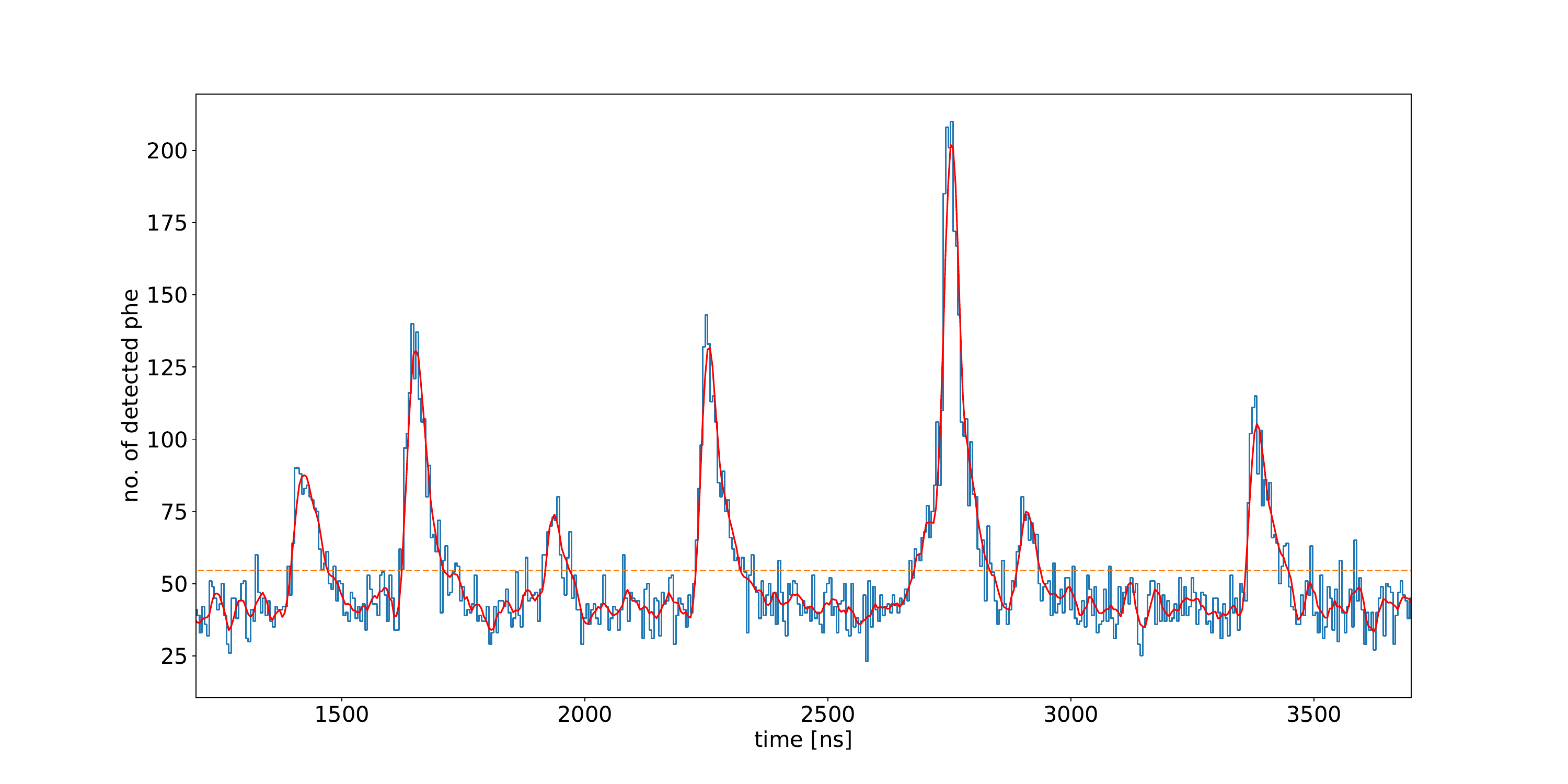}
\caption{Simulated waveform section illustrating the smoothing algorithm applied before the peak-finding routine (red line). The dashed horizontal line indicates the threshold above which a pulse is considered a candidate.}
\label{fig:smoothed_wf}
\end{figure}

\begin{figure}
\centering
\includegraphics[width=0.9\linewidth, trim={1.5cm 1.5cm 1.5cm 1.5cm},clip]{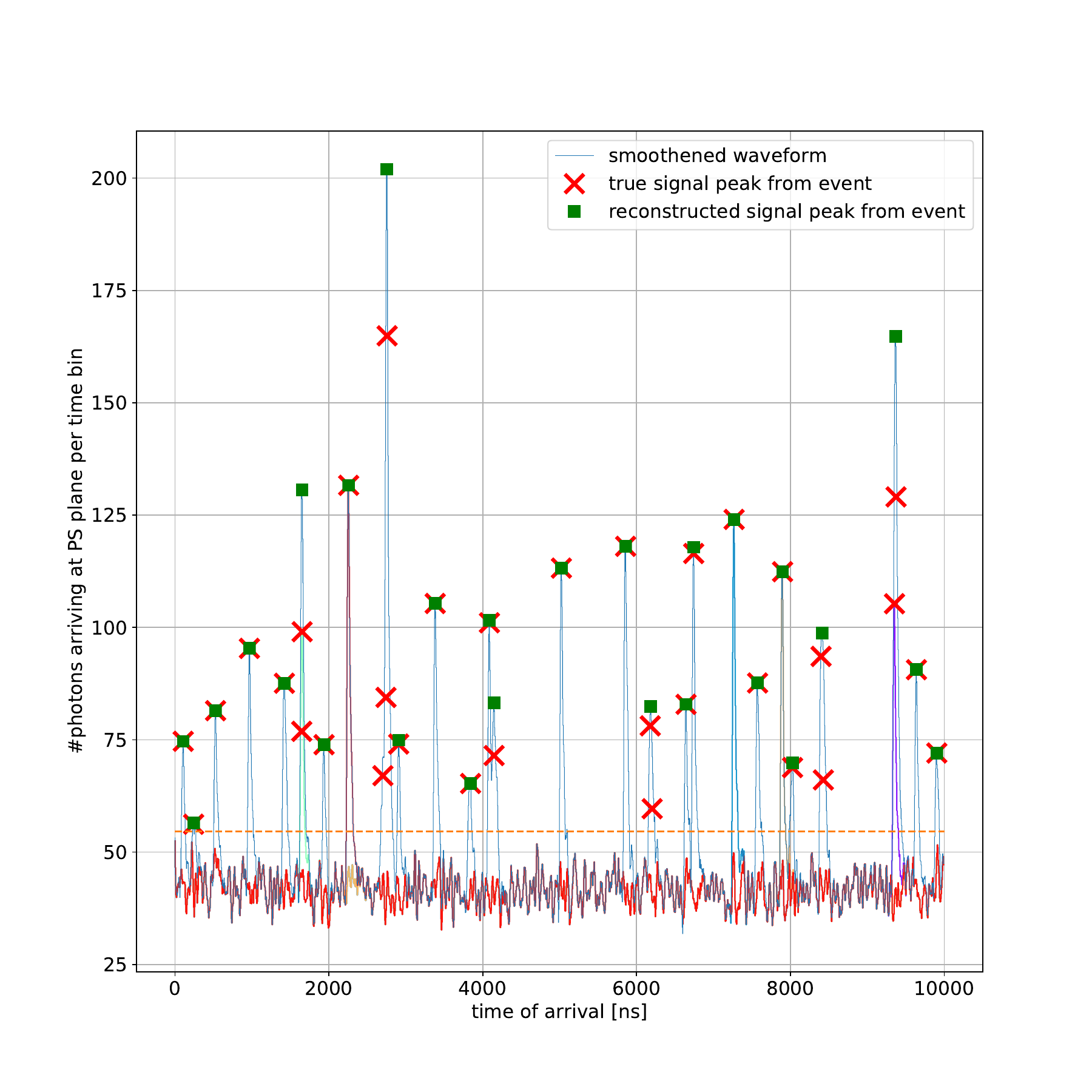}
\caption{Simulated TPC response over the entire photosensor plane for a muon multiplicity of 33 within a time window of $10~\mathrm{\mu{s}}$, considering configuration B (Teflon-lined TPC) at $-25$~\celsius. Muon peaks are represented with crosses (true) and squares (reconstructed), resulting in a peak-finding efficiency of 82\% in this example.}
\label{fig:s1peaks}
\end{figure}

Before the waveform is scanned for peaks, a Savitzky-Golay filter is applied to reduce noise coming from the dark count (figure~\ref{fig:smoothed_wf} shows a zoomed section of the waveform). Peaks are searched for through an algorithm that uses a set of conditions on the signal shape, such as peak height, prominence and width. The height also serves as the signal threshold, which has been set to $T = \lambda+2\cdot\sqrt{\lambda}\approx55$ phe (approximately 1~MeV), where $\lambda$ is the product of the average DCR per ns (at $-25$~\celsius) times the sampling bin, which has been set to 5~ns for this study. Furthermore, a peak candidate is required to have a prominence of 4.1~phe and a width of more than 2~ns, measured at 80\% relative distance to the peak. With these conditions, the purity of the peak-finding algorithm reaches 100\% well up to 100 muons per spill, with the efficiency remaining above 80\% for a multiplicity of $< 40$ muons (figure~\ref{fig:s1peaks}). These efficiency numbers are roughly compatible with a baseline occupancy estimated from the width of the scintillation pulses of about $40\times 0.025 /10 = 10\%$.

It is clear that, with the energy resolution values estimated previously for muons, the assignment of the reconstructed pulses to the ionization/track counterpart will become problematic, more so at high multiplicities.
One might anticipate the use of residual pulse-shape information or spatial correlations between the ionization and scintillation signals to perform the matching, as well as external time-tagging (if available), but the result of this procedure is setup-dependent and therefore beyond the scope of this study. Demonstrating the possibility of reaching the low-multiplicity threshold of 5~MeV for hadron reconstruction (figs.\ \ref{fig:energy_threshold} and \ref{fig:chi2_amplitude}) would require a similar analysis.

On the other hand, above 15~MeV, hadrons released in neutrino interactions inside the TPC will sit above this background field (at $2\sigma$), simplifying reconstruction as well as assignment, even at high multiplicities.

\section{Technical aspects}
\label{sec:technical}

\subsection{Photosensor response function and digitization}
\label{sec:Electronics}
The photosensor response function ($PRF$), signal-to-electronic-noise for single photons ($S/N_{elec}$) and sampling time ($\Delta{T}$) can modify the performance presented in the previous section. In order to address their impact, we consider light pulses from a canonical 5 MeV deposit in configuration B (Teflon-lined TPC) as before, this time at a position close to the cathode (blue line in fig.~\ref{fig:t0_vs_dcr}), to set the time resolution scale to around 1~ns. Extending over previous analysis, the $PRF$ is now introduced as a convolution over the Geant4-generated pulses, and the waveform template used for the fit includes the convolution with the $PRF$ as well. The decision to benchmark pulse-reconstruction through the timing performance is based on the observation that preserving time resolution also preserves energy resolution and energy thresholds.

We envisage SiPM ganging \cite{DIncecco:2017bau} in order to cover, in an affordable manner, an area of at least 7.5~m$^2$ (38\% coverage). The $PRF$ will depend on the parameters of the feedback and zero-pole cancellation loops of the amplifying electronics, and the equivalent SiPM capacitance at its input. We consider response functions that can be realistically obtained when ganging 16 channels, as per an ongoing design based on LTSpice, including the so-called Corsi parameters for model S14161-6050HS by Hamamatsu \cite{Corsi1, Corsi2}, following parallel passive ganging ($\times 4$) and active summing ($\times 4$). The 1'' photosensor footprint anticipates a lump sum of 12\,000 readout channels.

\begin{figure}
\centering
\includegraphics[scale=0.48]{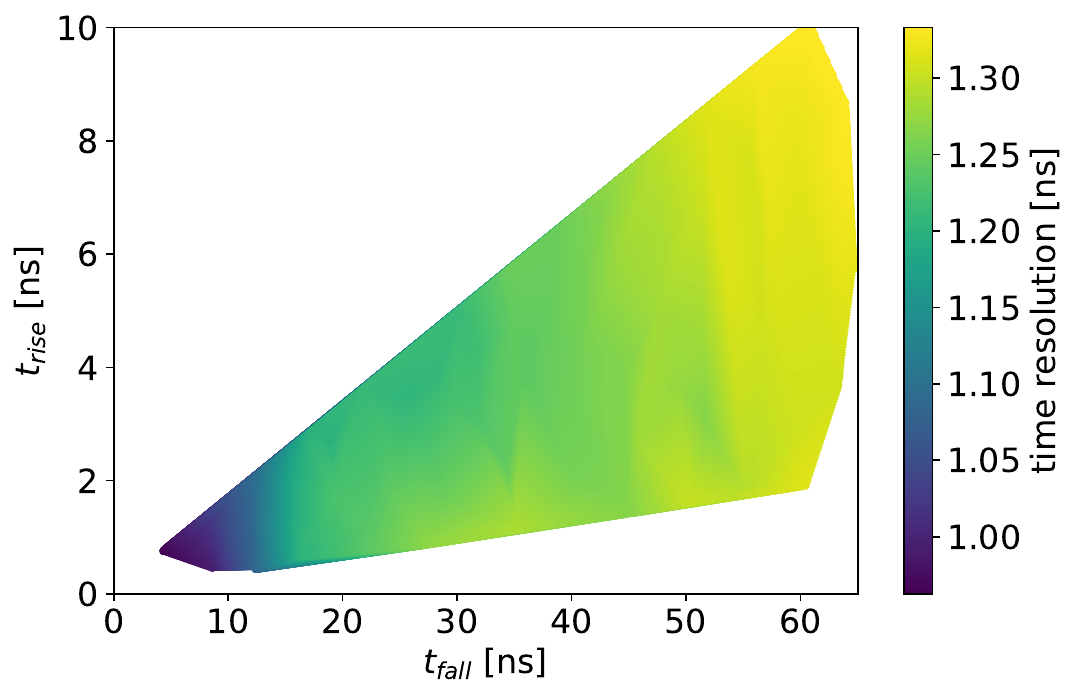}
\caption{Simulated time resolution for our reference case (5~MeV deposits) as a function of the $t_\mathrm{rise}$ and $t_\mathrm{fall}$ parameters of the photosensor response function ($PRF$), parameterized through a by-exponential function ($t_\mathrm{rise}$ and  $t_\mathrm{fall}$ are defined between 10\% and 90\% of the signal maximum).}
\label{fig:trisetfall}
\end{figure}
 
In simulation, a bi-exponential function of the type introduced in eq.~(\ref{Eqpulse}) allows a good description of the $PRF$, with two extreme cases being considered in the following: `fast', where $t_\mathrm{rise}=4.2$~ns and $t_\mathrm{fall}=13.8$~ns; and `slow', where $t_{rise}=10.1$~ns and  $t_{fall}=60.8$~ns (defined from 10\% to 90\%). For illustration, figure~\ref{fig:trisetfall} provides the time resolution over a range of $t_{rise}$, $t_{fall}$ values (sampled by Monte Carlo) including those two scenarios. Maximum variations stay well within a 30\%. The small effect observed underlines the fact that the scintillation response of Ar/CF$_4$, together with the delays due to light reflections in the TPC, dominate over the electronic response in the range of $PRF$s considered. 
 
\begin{figure}
     \centering
     \includegraphics[scale=0.48]{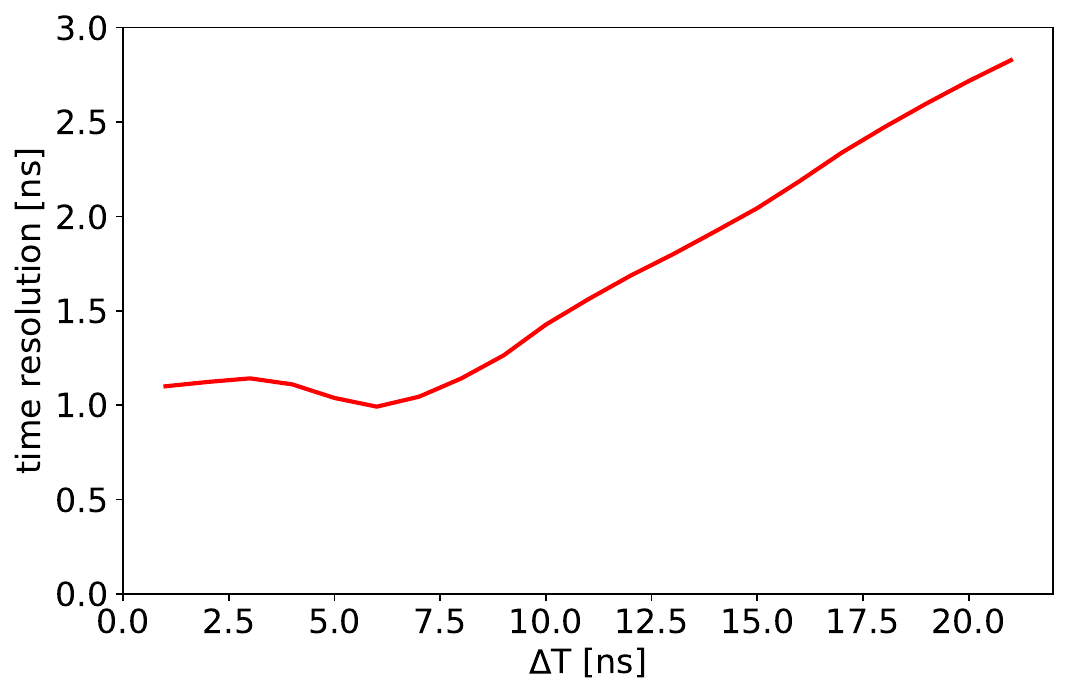}
     \caption{Simulated time resolution for our reference case (5~MeV deposits) as a function of the time binning employed during signal digitization (`fast' electronics response assumed).}
 \label{fig:sampling_time}
 \end{figure}

Another critical study is the determination of the optimum sampling time, which has a direct impact on the digitizing electronics, multiplexing strategy and cost. Again, as in previous section, the start time of the sampling process relative to the pulse has been randomized within a time bin, to mimic an asynchronous sampling. Figure~\ref{fig:sampling_time} shows that the time resolution is preserved for a time sampling corresponding to $\Delta T \leq$7~ns, that motivates the value of $\Delta T = 4$~ns chosen throughout the text.

\begin{figure}
\centering
\includegraphics[width=0.55\linewidth]{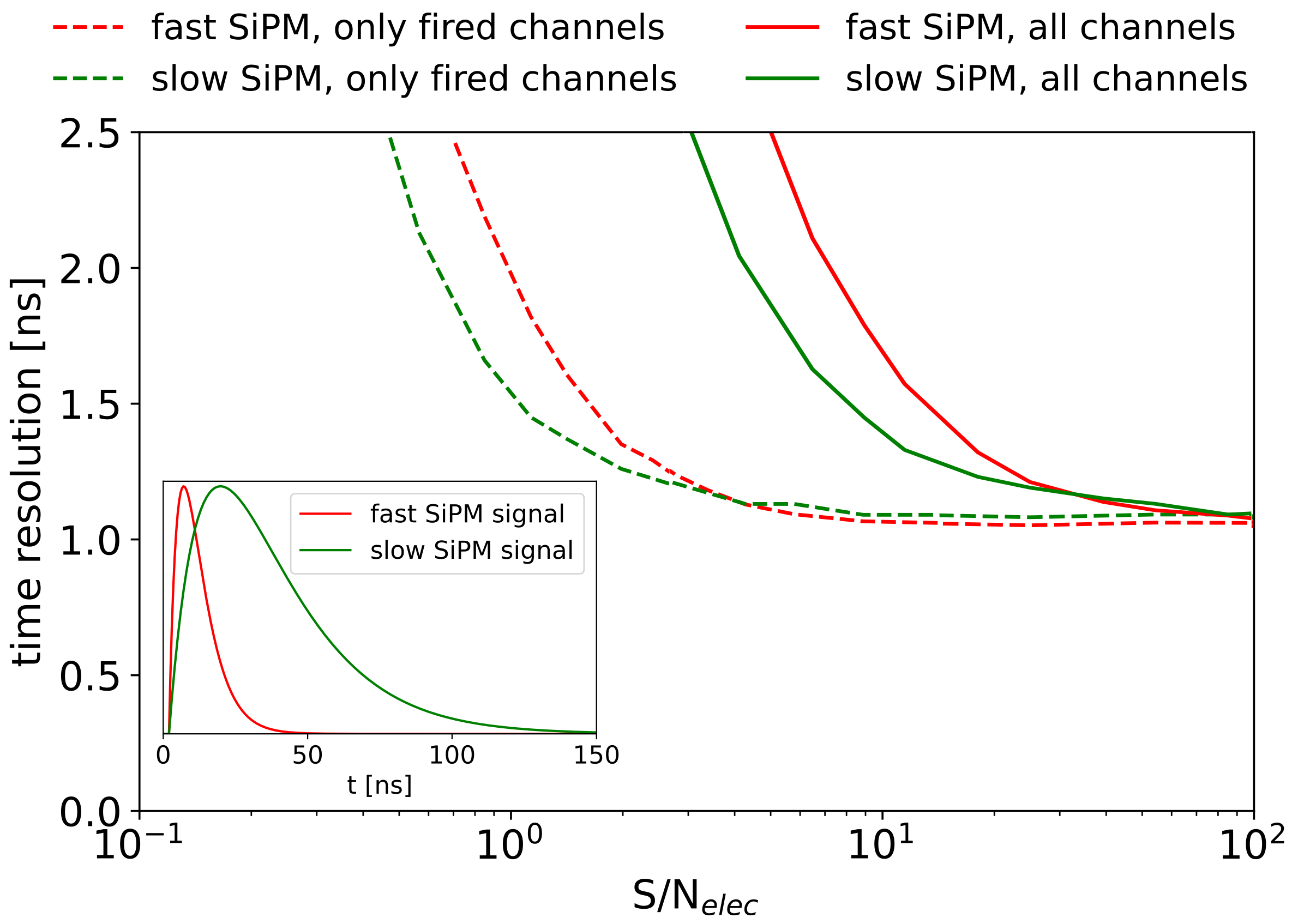}
\caption{Simulated time resolution for our reference case (5~MeV deposits) as a function of the amplitude of the signal over the electronic noise ($S/N_{elec}$) for single photons. The inset shows the two electronic responses studied (`fast' and `slow').}
\label{fig:amplifier_noise}
\end{figure}

Finally, we evaluate the impact of the electronic noise in the pulse reconstruction. We assumed a white-noise spectrum with a normal amplitude distribution centered around 0, whose $\sigma$-value is referred to as $N_{elec}$. The simulated pulses are reconstructed after smearing the amplitude in each 4~ns time-bin, following this noise model. The time resolution as a function of the ratio of the signal amplitude to noise ($S/N_{elec}$) is shown in figure~\ref{fig:amplifier_noise} (the $S/N_{elec}$ refers to that of the single-photon response). With continuous lines we present the performance when adding all electronic channels (totalling 12000 for the fill factor of $38\%$ discussed in text, 16-fold ganging, and assuming our 1 squared-inch baseline S14161-6050HS SiPM-sensor). With discontinuous lines we show the result of adding only channels that fired at a particular time.

\begin{figure}
\centering
\includegraphics[width=0.495\linewidth, trim={0 0 0 10cm},clip]{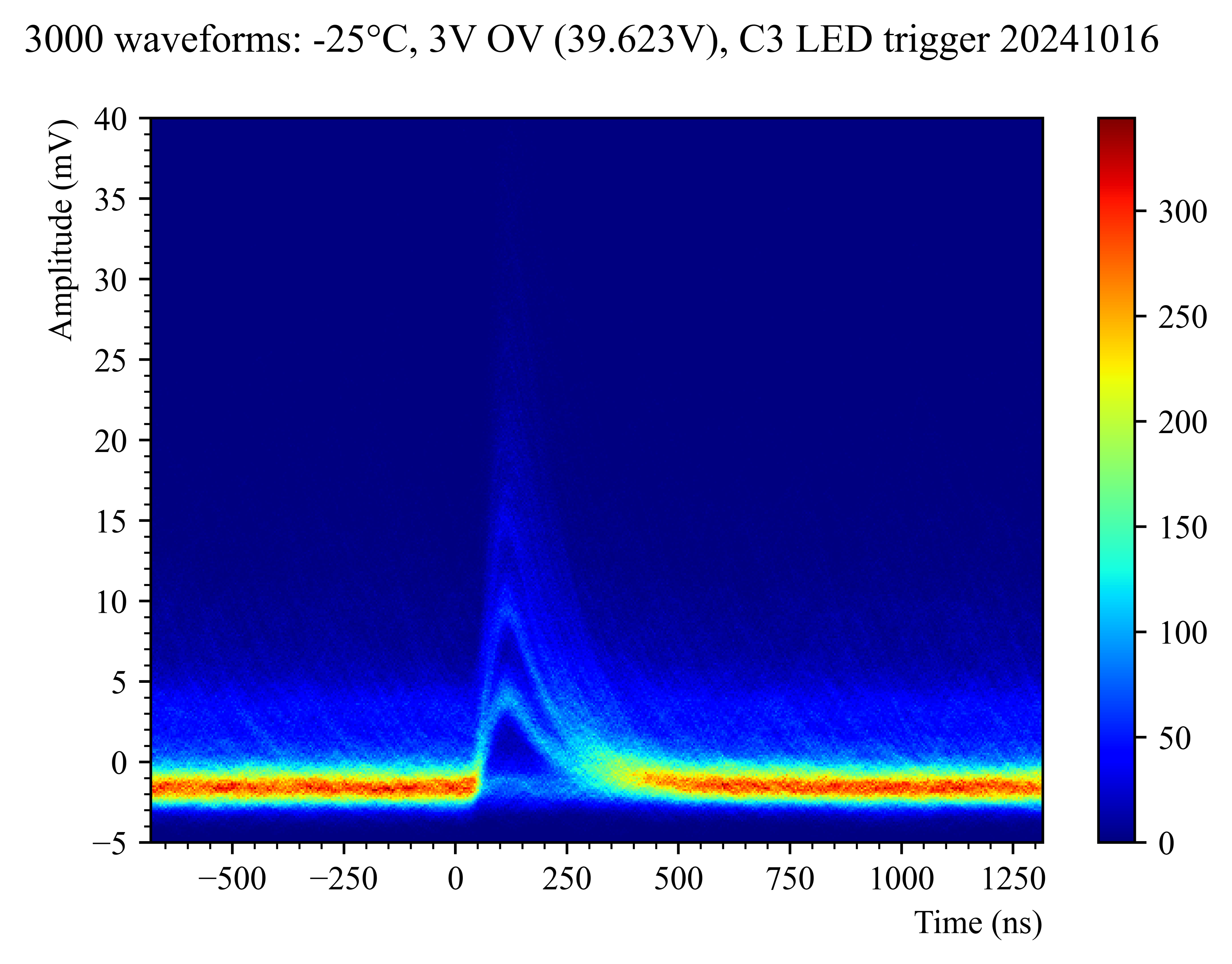}
\includegraphics[width=0.495\linewidth, trim={0 0 0 5cm},clip]{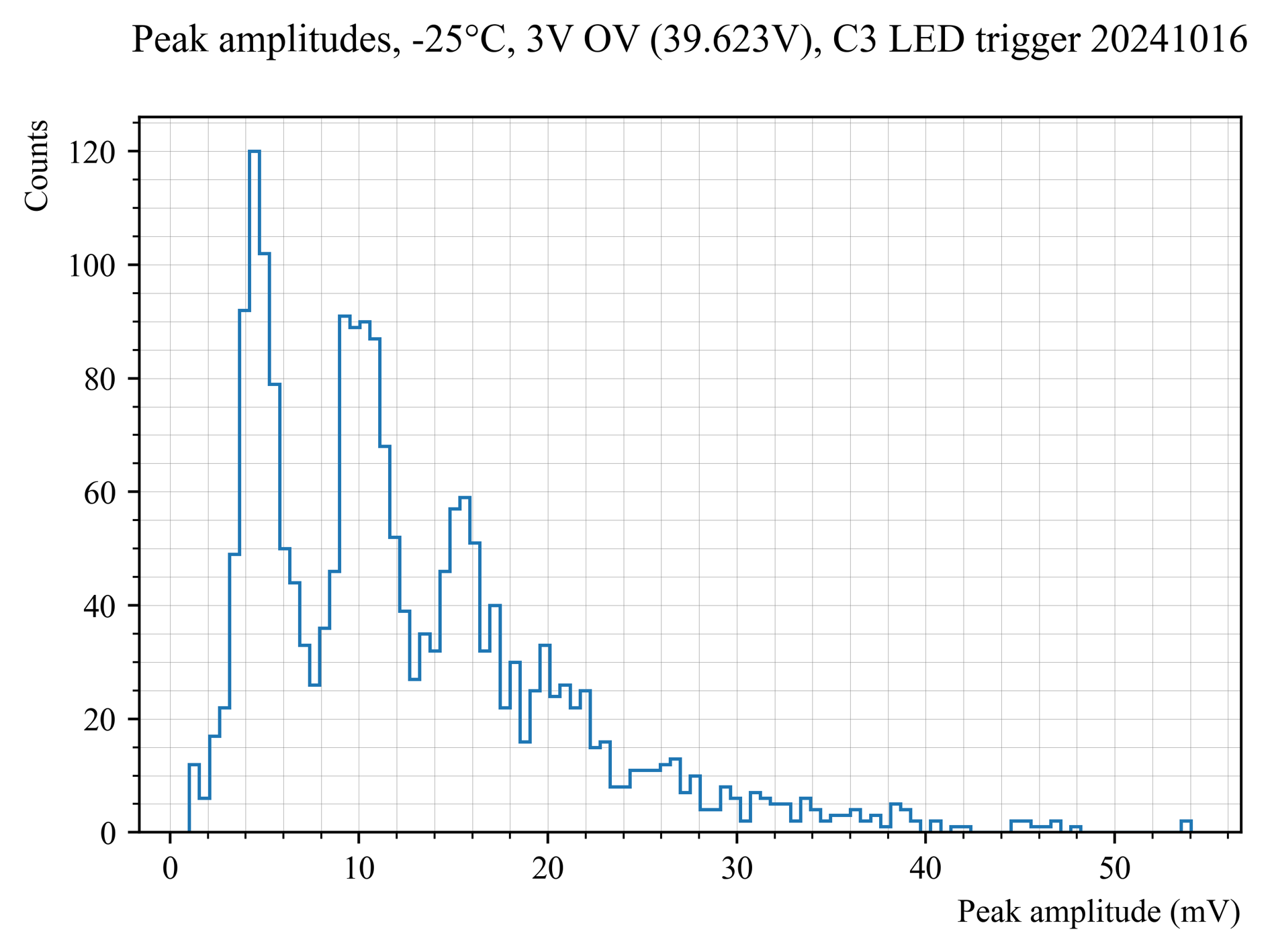}
\caption{Left: oscillogram taken in persistency mode, showing preliminary results for SiPM photon peaks at $T=-25~\celsius$, obtained in a development board reading out 16 signals from a S14161-6050HS 1'' SiPM sensor. The width of the $PRF$ and $S/N_{elec}$ is in the range of those discussed in-text. Right: spectrum obtained from the peak amplitudes of the pulses shown in the left panel.}
\label{fig:sipms_electronics}
\end{figure}

As the dark rate occupancy in a 100~ns window is only about 6\%/channel at $T=-25$~\celsius, single-photon reconstruction should remain largely unbiased, and a threshold may be in principle applied to every channel to select only those that fired in the given time window, if signal-to-noise allows. A signal-to-noise ratio around 5--10 should easily allow single-photon detection with a threshold of $3$--$5\sigma$ above the noise level and, according to fig.~\ref{fig:amplifier_noise}, it should also keep the pulse-reconstruction (time resolution) of 5~MeV deposits largely unaffected (dashed lines). This seems well within reach of current technology as signal-to-noise ratios above 10 have been reported on areas similar to the ones discussed here \cite{DIncecco:2017bau}. In case of adding all channels, on the other hand, a signal-to-noise value as high as 50 would be in principle needed to maintain good reconstruction down to 5~MeV energies (continuous lines in fig.~\ref{fig:amplifier_noise}), that seems extremely challenging. Clearly, any form of common-mode noise should be suppressed, and differential or pseudo-differential readouts seem a must at the proposed scale. Figure~\ref{fig:sipms_electronics} shows, for illustration, results obtained for the photo-sensor type discussed in text, in a development board ganging its 16 channels into a single output. An oscillogram of the dark rate pulses (fig.~\ref{fig:sipms_electronics}, left panel) in persistency mode is presented, with the single, double and triple photon peaks being clearly visible. Only one of the two legs of the pseudo-differential outputs of the development board has been read out. The right panel in fig.~\ref{fig:sipms_electronics} is a histogram of the amplitude values, showing a $S/N_{elec}$ above 5.

\subsection{Light collectors}
Even for an isotropic point-like light source, a hollow cylindrical reflector causes a bias in the landing angles on its endcaps, towards perpendicular incidence. The further the source is from the endcaps, the more likely a photon will land close to the normal. This observation opens the possibility to use conventional light collectors to reduce the active area of the photocathode, potentially increasing the signal-to-DCR ratio and reducing cost.

A light collector would ideally need to be optimized for the most relevant positions of an event within the chamber. However, in a neutrino TPC that is expected to operate under highly homogeneous neutrino and muon fluxes, the required depth-of-field of any light collection system coincides with the TPC itself. Thus, the following simplifying assumptions have been made: (i) all light collectors are assumed to be identical; (ii) as the angular distribution of the landing photons was found to be rather independent from the radial position of the source, we consider photon sources distributed homogeneously over disks placed at different $z$ distances from the photocathode. As usual, simulations in configuration B (Teflon-lined TPC) were chosen for reference. 

\begin{figure}
\centering
\includegraphics[width=0.6\linewidth]{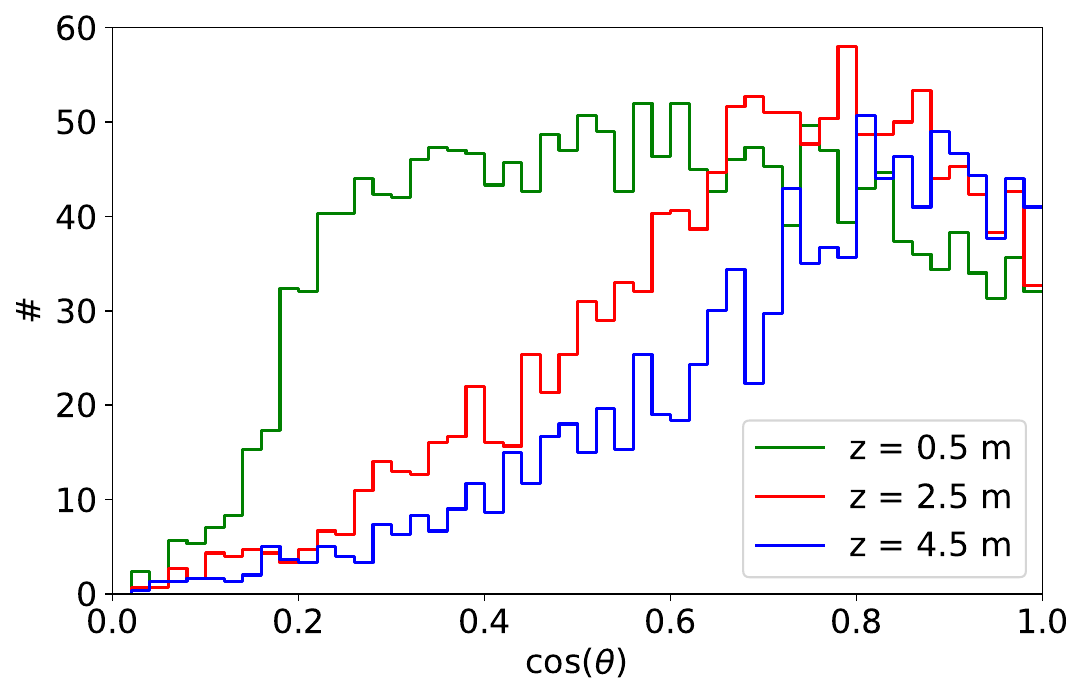}
\caption{Distribution of photon landing angles at the photosensor plane of the TPC, for light sources placed at different distances from the photocathode (configuration B: Teflon-lined TPC): $z=0.5$~m (green), $z=2.5$~m (red), $z=4.5$~m (blue).}
\label{fig:angular distribution}
\end{figure} 

In the above conditions, the aforementioned focusing effect is apparent in figure~\ref{fig:angular distribution}, as events far from the photocathode show a relatively narrow distribution close to perpendicular incidence ($\theta=0^\circ$) while the distribution of events closer to the photocathode resembles isotropic emission down to some cutoff angle (coming from the photocathode size). This is indeed an appealing situation as it can be anticipated that the light collector at the photosensor plane will perform at its best when the light collection of the TPC is lower, and vice-versa.

We choose a classical Winston cone as an illustrative example, since any other non-imaging optics is expected to follow closely the conclusions obtained for it. The critical angle $\Theta_c$ defines the cone geometry as \cite{WinCone}:
\begin{equation}
    \frac{A_{w}}{A_{ps}} = \left(\frac{n}{\sin{\Theta_c}} \right)^2 \,,
    \label{W-cone} 
\end{equation}
with $A_{w}$ referring to the area of the cone's front face (`window'), $A_{ps}$ referring to the area of the cone's rear face (`photosensor') and $n$ being the refraction index of the cone's material. 
A good compromise between the light removed by the angular cut, the size of the Winston cone and the reduction in photosensor area seems to exist at around $\Theta_c=60^{\circ}$. We implement this condition through a PMMA-based Winston cone ($n\simeq 1.5$), yielding a ratio $A_w$/$A_{ps}$ of 3. If aiming at $\sqrt{ A_{ps}} \sim 1''$ (as in the concept proposed here, based on the S14161-6050HS SiPM model), a critical angle of $\Theta_c=60^{\circ}$ implies that the length of the PMMA cone would be about 20~cm (and 12\,000 such cones would be necessary).

\begin{figure}
\centering
\includegraphics[width=0.6\linewidth]{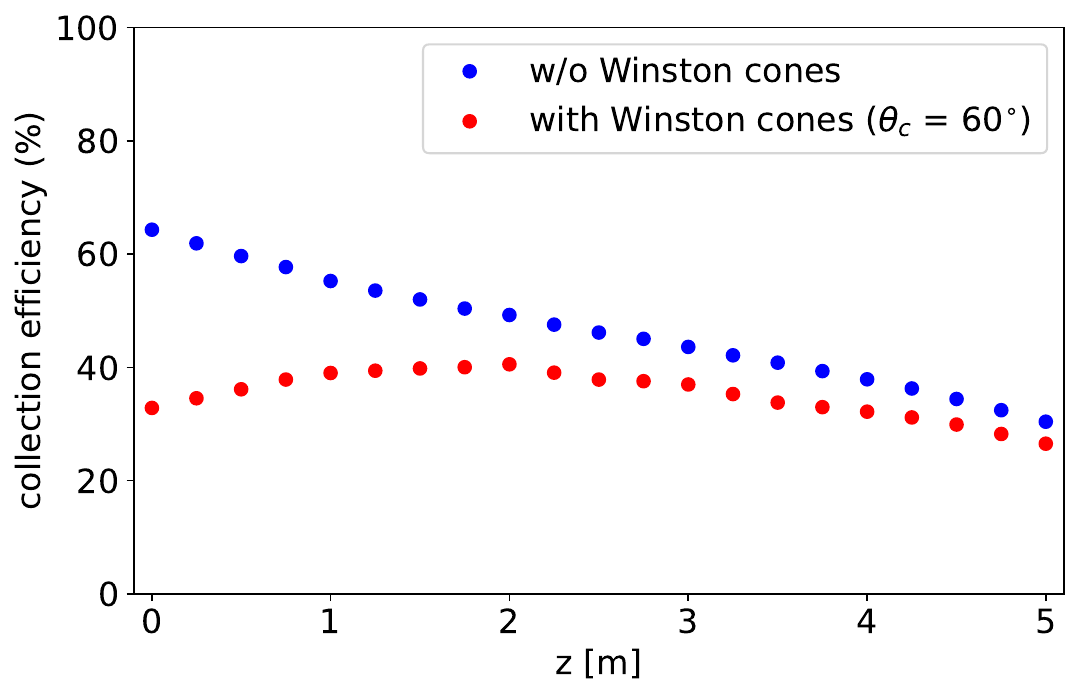}
\caption{Light collection efficiency of the TPC (configuration B: Teflon-lined), as a function of the distance of the source to the photosensor plane. Results are presented with (red data points) and without (blue data points) the use of PMMA Winston cones, for a critical angle of $\Theta_c=60^{\circ}$ (area reduction equals $\times 3$). Red circles indicate the best case, as transmission and reflection losses have been neglected.}
\label{fig:eff_WinstonCones}
\end{figure}

The total light collection efficiency of the TPC (configuration B) with and without Winston cones is shown in figure~\ref{fig:eff_WinstonCones} as a function of the distance of the source to the photosensor plane, in the ideal limit where transmission and reflection losses in the cone are neglected. As anticipated, the use of Winston cones makes the light collection much flatter overall, at the 30--40\% level (red data points). Although, in view of the potential reduction of the active area of the photocathode ($\times 3$ in this case) this result is encouraging, it must be certainly taken with a grain of salt: the actual performance of such a light collector in terms of light transmission and reflectivity will require experimental verification. Additional gains are expected, on the other hand, for materials with higher refraction index such as sapphire.

\subsection{Photosensor cooling}
\begin{figure}
\centering
\includegraphics[width=0.75\linewidth]{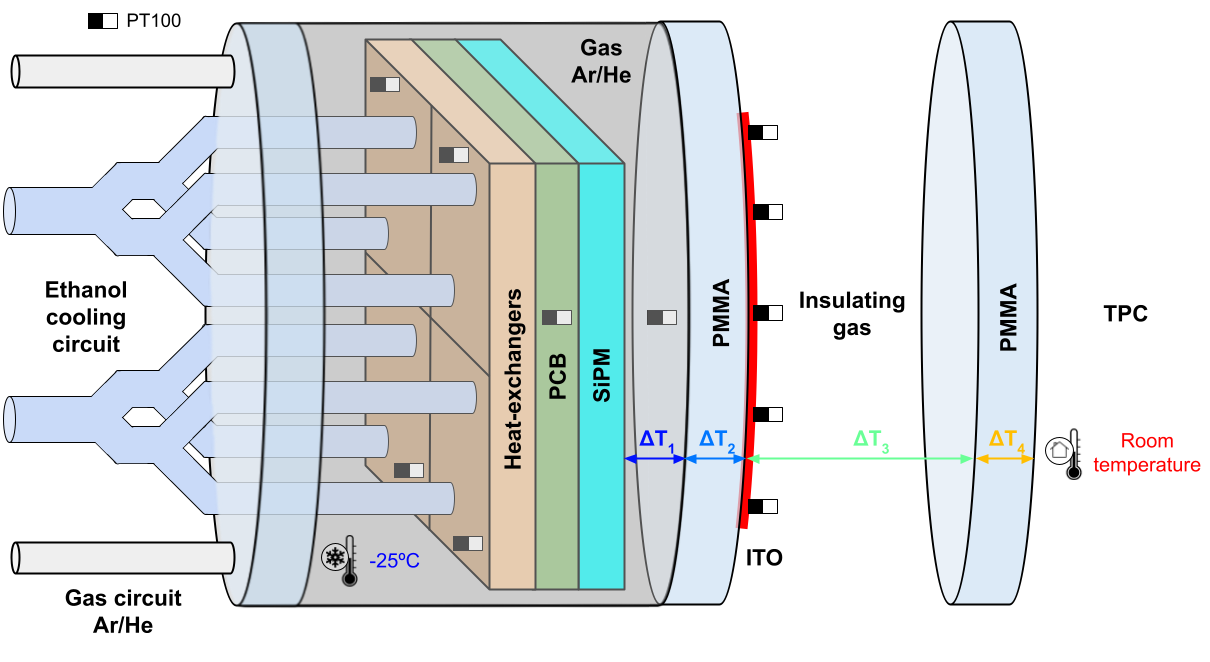}
\includegraphics[width=0.75\linewidth]{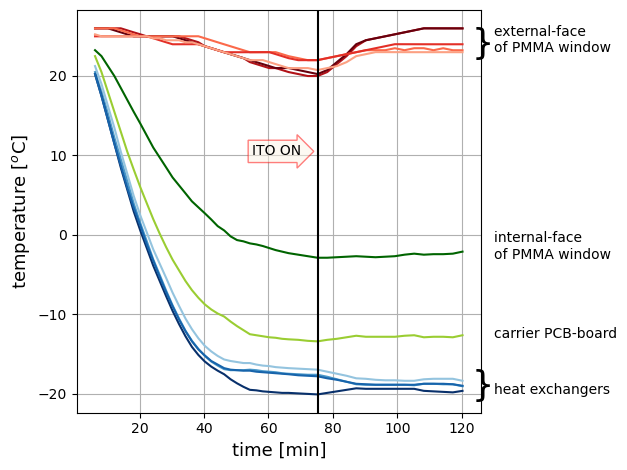}
\caption{Top: a possible concept for a non-vacuum-based cryostat, relying on passive insulation and a mild external heating worth 100~W/m$^2$. Bottom: experimental results, illustrating the effect of the external heating at stabilizing the external temperature. Heating was activated at the time marked with the arrow `ITO on'. The carrier board had neither through-vias nor thermal pads in this test.}
\label{fig:cool}
\end{figure}

A temperature of $-25$~\celsius\ is not alien to the operation of silicon-based photosensors \cite{PANDA:2008rpr} and, with due precautions, it is high enough so as not to expect strong tin pest effects on the auxiliary electronic boards. Although several cooling strategies are possible, we discuss briefly a simple implementation that avoids the use of large vacuum vessels, and that might be practical when targeting operation near a pressurized system, at a modest power consumption. Given the necessity to avoid temperature gradients in the TPC, a combination of passive insulation and a mild active heating of the external window surface is proposed. Passive insulation could be enabled by 5~mm of pressurized Ar gas and a 20~mm thick PMMA window, coated with an indium-tin-oxide (ITO) conductive film on its external surface, as shown in figure~\ref{fig:cool} (top). The presence of a second PMMA window at the cathode plane ensures a buffer gas region for homogenization of residual temperature gradients over the windows. Fig.~\ref{fig:cool} (bottom) shows the experimental results for a 10~cm diameter cryostat designed according to these principles and cooled down to $-20$~\celsius\ through an ethanol chiller. Upon applying a voltage across the ITO film ($\simeq 12$~V), marked with an arrow as `ITO on', the internal cooling power can be balanced and the system returned to stationary conditions, with a power consumption of about 100~W/m$^2$. Over the entire photosensor region, this would amount to a modest 750~W (to which the electronics power needs to be added).

\section{Summary of main results and conclusions}
\label{conclu}
This work presents a feasibility analysis of the reconstructibility of the primary scintillation signal in a pressurized TPC based on Ar/CF$_4$ (99/1), with the capacity to house up to 1.5~tons of argon (5~m height, 5~m diameter) at 10~bar. One of the main motivations for such a system is its ability to accurately reconstruct low-energy hadrons emerging from neutrino interactions that cannot be well-resolved by conventional means, either in condensed phase (tracks too short) or in pure gas phase (tracks long enough to be resolved, but not long enough to allow external time-tagging), thus biasing the neutrino-energy estimate. A second motivation is technical: to increase the stand-alone tracking and timing capabilities of the TPC, providing independent means of disambiguating the time and $z$-position of the interactions. A third motivation, which has yet to be demonstrated at high pressures and will be discussed elsewhere, is the potential to achieve high-resolution full-3D optical imaging of the particle tracks.

With this in mind, part of our study focused on the reconstruction of the primary scintillation released by 5 MeV protons, which provides a reasonable proxy for the shortest track that can be reconstructed at the anode (tracking) plane under the discussed gas conditions (see, e.g., \cite{DUNE:2021tad}). We conclude that, upon lining the TPC field cage with PTFE/Teflon: (i) neutral-current neutrino interactions with visible energy in the range of 5--200 MeV, (ii) charged-current neutrino interactions where the associated lepton would not be time-tagged externally, or (iii) stranded neutrons scattering off protons in the gas and transferring more than 5 MeV, would show a significant scintillation counterpart. This allows for spill and vertex assignment, and ns-level time-of-flight determination. In general, any low-energy process (e.g., beyond the Standard Model phenomena such as milli-charged particles) could similarly benefit from the TPC's ability to perform stand-alone time-tagging.

Additional technical requirements to achieve the aforementioned performance include: (i) a photosensor coverage of the cathode plane of at least 38\% (7.5~m$^2$), (ii) use of a SiPM-array, (iii) operation at $-25~\celsius$, and (iv) read-out grouped by at least a factor of 16. Despite the general academic interest in neutrino generators for understanding hadro-production down to nuclear binding energies, current knowledge of neutrino-nucleus interactions suggests a more modest energy scale, around 20~MeV, as being of immediate interest. Two physical cases identified in ref.~\cite{DUNE:2022yni} are: (i) reconstruction of the energy spectrum of protons modified due to final state interactions \cite{Theo1, Theo2, Theo3}, and (ii) multi-pion production \cite{Theo4}. Given this, and considering the evolving understanding of neutrino interactions with nuclei, we briefly discuss the response to a 20~MeV hadron. In such a scenario, the Teflon reflector around the field cage might not be necessary, as this energy would be above threshold for almost the entire chamber, with a time resolution below 2~ns. With Teflon lining, a modest 10\% photosensor coverage would suffice to provide similar performance. If maintaining the Teflon lining and the proposed 38\% coverage, operation of the photosensor plane close to $0~\celsius$  would be possible. Approaching an $\mathcal{O}$(MeV) threshold, however, will certainly require all three assets (coverage, lining, and cooling) simultaneously.

Finally, given the importance of reconstructing leptons from charged-current interactions for neutrino-oscillation experiments, we summarize the detector performance for `internal' and `external' muons. Although for neutrino energies at the GeV, lepton time-tagging could be done with external detectors in most cases, our results support that a time-of-flight measurement of these (or any other minimum ionizing track) in the 1--2~ns range over a few meters is possible. However, our work also shows that reconstructing and properly assigning primary scintillation pulses will become problematic for background multiplicities around a few tens of muons per (10~\textmu s) spill. Hadrons with energies above 15~MeV will be significantly above this muon field in terms of energy loss and can be expected to represent a realistic reconstruction threshold for high background multiplicities.

Our work also provides a plausible ---\thinspace yet largely conceptual\thinspace--- technical path towards a practical implementation of the proposed photosensor system, involving the possible use of SiPM ganging, active cooling, and Winston cones. Preliminary engineering results supporting this idea have been presented.

\bmhead{Acknowledgements}
This research has received financial support from the European Union’s Horizon 2020 Research and Innovation programme under GA no.\ 101004761, from Xunta de Galicia (Centro singular de investigación de Galicia accreditation 2019-2022), from the Generalitat Valenciana (grant CIDEGENT/2019/049), and by the “María de Maeztu” Units of Excellence program MDM-2016-0692. DGD was supported by the Ram\'on y Cajal program (Spain) under contract number RYC-2015-18820. JM-A acknowledges support from the Ram\'on y Cajal program (grant RYC2021-033265-I) funded by the Spanish MCIN/AEI/10.13039/501100011033 and by the EU (NextGenerationEU/PRTR). This research was also partly funded by the Spanish Ministry (‘Proyectos de Generación de Conocimiento’, PID2021-125028OB-C21).

Special thanks must be given to Alan Bross, Carlos Escobar and Adam Para (Fermilab) for encouragement and many insightful discussions.

\bibliography{references}

\end{document}